\documentclass[twocolumn,prb,epsf,superscriptaddress]{revtex4-2}
\usepackage{amsmath}
\usepackage{amsfonts}
\usepackage{epsfig}
\usepackage{epsf}
\usepackage{array}
\usepackage{color}
\usepackage{ulem}
\usepackage{multirow}
\usepackage{upgreek}

\begin{document}

\title{Topological flat bands in a kagome lattice multiorbital system}
\author{Satoshi Okamoto}
\altaffiliation{okapon@ornl.gov}
\affiliation{Materials Science and Technology Division, Oak Ridge National Laboratory, Oak Ridge, Tennessee 37831, USA}
\author{Narayan Mohanta}
\affiliation{Materials Science and Technology Division, Oak Ridge National Laboratory, Oak Ridge, Tennessee 37831, USA}
\author{Elbio Dagotto}
\affiliation{Materials Science and Technology Division, Oak Ridge National Laboratory, Oak Ridge, Tennessee 37831, USA}
\affiliation{Department of Physics and Astronomy, The University of Tennessee, Knoxville, Tennessee 37996, USA}
\author{D. N. Sheng}
\affiliation{Department of Physics and Astronomy, California State University, Northridge, California 91330, USA}

\maketitle


\section*{Abstract}
Flat bands and dispersive Dirac bands are known to coexist in the electronic bands in a two-dimensional kagome lattice. 
Including the relativistic spin-orbit coupling, such systems often exhibit nontrivial band topology, 
allowing for gapless edge modes between flat bands at several locations in the band structure, 
and dispersive bands or at the Dirac band crossing. 
Here, we theoretically demonstrate that a multiorbital system on a kagome lattice is a versatile platform to explore  the interplay between 
nontrivial band topology and electronic interaction. 
Specifically, here we report that the multiorbital kagome model with the atomic spin-orbit coupling naturally supports topological bands 
characterized by nonzero Chern numbers $\cal C$, 
including a flat band with $|{\cal C}| =1$. 
When such a flat band is $1/3$ filled, the non-local repulsive interactions induce a fractional Chern insulating state. 
We also discuss the possible realization of our findings in real kagome materials. 





\section*{Introduction}
%
Flat-band systems have been proposed as interesting theoretical models to prove the existence of ferromagnetic ordering with itinerant electrons
\cite{Lieb1989,Mielke1991,Tasaki1992,Mielke1993}. 
Theoretical developments in such flat-band systems have been made almost in parallel with those in the widely discussed topological insulators (TIs) 
\cite{Thouless1982,Haldane1988,Kane2005,Bernavig2006}. 
The nontrivial topology of electronic bands in a kagome lattice, one of those flat-band systems, has been extensively studied 
\cite{Ohgushi2000,Guo2009,Wen2010,Liu2013,Kiesel2013,Mazin2014,Zhou2014,Xu2015a,Yamada2016,Bolens2019}. 

The experimental quests for topological materials with kagome lattice have also been carried out.  
Many of such experimental efforts were stimulated by the prediction of Weyl semimetals \cite{Wan2011,Xu2015b}, 
including 
intermetallic compounds 
involving Co \cite{Alled2012,Yin2019,Jiao2019,Liu2019,Meier2020}, 
Fe \cite{Ye2018,Lin2018,Kan2020,Sales2019}, 
Mn \cite{Nakatsuji2015,Kuroda2017,Nayak2016}, 
and van-der-Waals compounds \cite{Park2020}, 
as well as optical lattices \cite{Taie2015,Drost2017}. 
More recently, the coexistence of superconductivity and nontrivial band topology was reported 
in a kagome compound \cite{Ortiz2020,Wu2021,Feng2021,Denner2021}. 

When a flat band is partially occupied by electrons, the Coulomb repulsive interactions could become dominant over the electronic kinetic energy. 
This situation is already realized in two-dimensional electron gases under applied magnetic fields, where flat bands correspond to Landau levels.  
Fractional quantum Hall (FQH) effects were thus discovered \cite{Tsui1982,Laughlin1983}. 
An exact numerical analysis made an important contribution by demonstrating that quantum fluctuations are essential to stabilize FQH states over charge density wave states \cite{Yoshioka1983}. 
Once the charge excitation gap is induced at a fractional filling, the property of FQH states is elegantly explained using effective theory \cite{Jain1989}.

Recently, further intriguing proposals were put forward by considering flat bands with nontrivial topology and repulsive interactions, 
whereby FQH states could be generated without having Landau levels, called fractional Chern insulators (FCIs). 
These proposals considered single-band models on 
a kagome lattice \cite{Tang2011,Wu2012a}, checkerboard lattices \cite{Sun2011,Neupert2011,Sheng2011,Wang2011,Regnault2011}, a Haldane model on a honeycomb lattice and a ruby lattice \cite{Wu2012a}, 
as well as 
multi-band models on a buckled honeycomb lattice \cite{Xiao2011}, a triangular lattice \cite{Venderbos2012}, and 
a square lattice for the mercury-telluride TI \cite{Wu2012a}. 
It was later revealed that quantum Hall states realized in flat band systems and 
those realized under an applied magnetic field are adiabatically connected \cite{Wu2012b}. 
When realized in real materials, FCI states in flat band systems could become a vital element of topological quantum computing \cite{Moore1991,Nayak2008}. 
Based on numerical results \cite{Tang2011,Wu2012a,Sun2011,Neupert2011,Sheng2011,Wang2011,Regnault2011,Xiao2011,Venderbos2012}, 
the possibility of FCI states was suggested in some flat band systems \cite{Liu2013,Zhou2014,Yamada2016}.  
However, the material realization of such FCI states has yet to be demonstrated as 
theoretical proposals often focus on simple one-band models and other proposed systems have small band gaps. 

Motivated by the recent experimental realization of kagome materials, 
where multiple transition-metal $d$ orbitals are active near the Fermi level, 
we consider in this work a multiorbital itinerant-electron model on a two-dimensional kagome lattice. 
With the atomic spin-orbit coupling (SOC), this model shows multiple topological phases, 
including spin Hall insulators when spin splitting is absent and Chern insulators when spin splitting is induced. 
Furthermore, this model exhibits flat bands having nonzero Chern number as in a single-band kagome system \cite{Guo2009}. 
We found that non-local Coulomb interactions induce FCI states when such a flat band is fractionally occupied by electrons. 
Note that our approach employs the original on-site local source of the SOC, 
while most of simplified models widely employed in other efforts assume the form of SOC simply based on symmetry considerations. 
Thus, our work relies on more fundamental foundations. 
Our model calculation is particularly relevant to CoSn-type intermetallic compounds 
when a single kagome layer becomes available. 

\section*{Results}
\begin{figure}
\begin{center}
\includegraphics[width=0.9\columnwidth, clip]{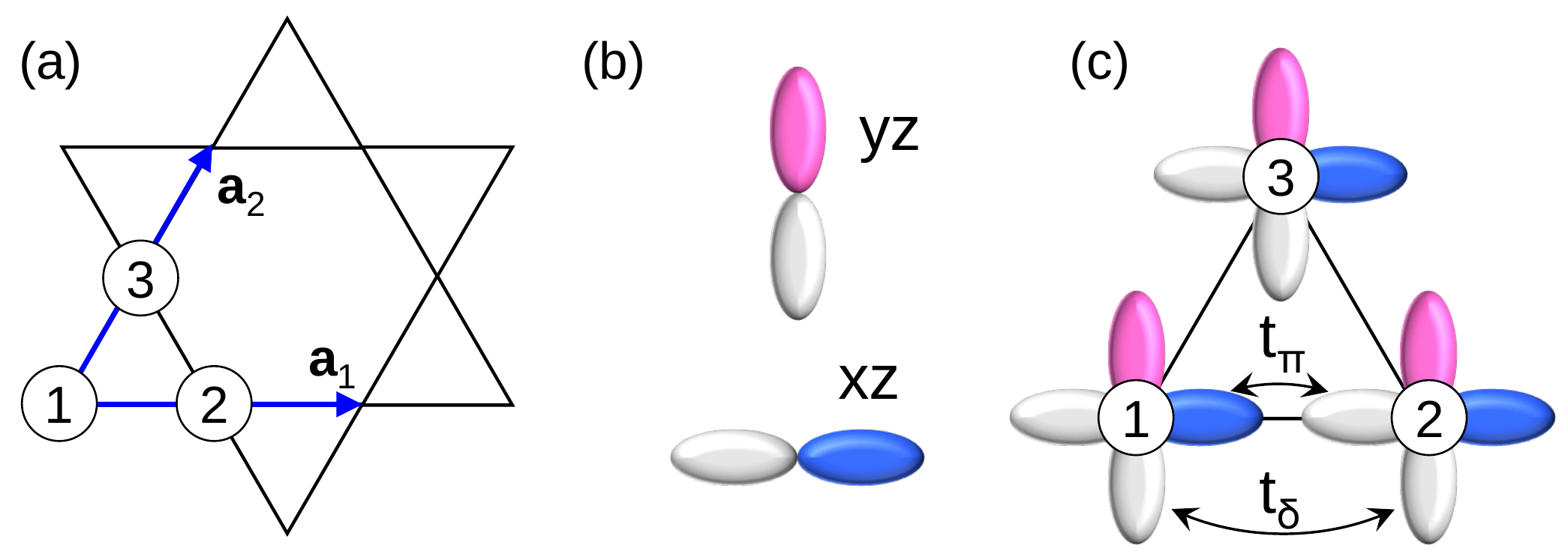}
\caption{Schematics of our theoretical model. 
(a) Kagome lattice with three sublattices, labeled 1, 2, and 3. 
The two arrows are lattice translation vectors ${\rm \bf a}_{1,2}$. 
(b) Local orbitals $a=yz$ and $b=xz$. 
Colored ellipsoids indicate regions of electron wave functions, where the sign is positive. 
(c) Nearest-neighbor hopping integrals. 
$yz (xz)$ orbitals between site 1 and site 2 are hybridized via diagonal hopping $t_{\updelta (\uppi)}$, i.e., $\updelta (\uppi)$ bonding. 
Other hopping integrals between site 2 and site 3 and  between site 1 and site 3 are obtained via the Slater rule \cite{Slater1954} 
as shown in the Method section. 
}
\label{fig:cartoon}
\end{center}
\end{figure}

{\flushleft \bf Theoretical model.}
%
To begin with, we set up a multi-orbital tight-binding model on a kagome lattice 
\begin{eqnarray}
H_{\rm t} =
- \sum_{\langle \rm{\bf r} \, \rm{\bf r}' \rangle} \sum_{\alpha \, \beta \, \sigma}
 \Bigl( t_{\rm{\bf r} \, \rm{\bf r}'}^{\alpha \beta} c_{\rm{\bf r} \alpha \sigma}^\dag c_{\rm{\bf r}' \beta \sigma} + {\rm H.c.} \Bigr) , 
\label{eq:hopping}
\end{eqnarray}
as schematically shown in Fig.~\ref{fig:cartoon}. 
Here, $c_{\rm{\bf r} \alpha \sigma}^{(\dag)}$ is the annihilation (creation) operator of an electron at site $\rm \bf r$, 
orbital $\alpha$, and with spin $\sigma= \uparrow$ or $\downarrow$. 
As discussed by Meier {\it et al.}\cite{Meier2020}, CoSn-type kagome systems have several flat bands 
with $\{yz,xz\}$, $\{xy,x^2-y^2\}$, or $3z^3-r^2$ character. 
We focus on a $\{yz,xz\}$ subset for simplicity and use $\alpha=a$ for the $yz$ orbital and $b$ for the $xz$ orbital. 
With this basis, 
nearest-neighbor hopping intensities $t_{\rm{\bf r} \, \rm{\bf r}'}^{\alpha \beta}$ can be parameterized using Slater integrals\cite{Slater1954}.
Between site $\bf 1$ and site $\bf 2$, $\hat t_{\bf 1 \,  2}$ is diagonal in orbital indices as $t_{\bf 1 \, 2}^{aa} = t_\updelta$ and $t_{\bf 1 \, 2}^{bb} = t_\uppi$, 
corresponding to $(dd\updelta)$ and $(dd\uppi)$, respectively, by Slater and Koster\cite{Slater1954}. 
Other components are obtained by rotating the basis $a$ and $b$ as shown in the Methods section. 
From now on, $t_\uppi$ is used as the unit of energy.

Because $yz$ and $xz$ are written using the eigenfunctions of angular momentum $l_z=\pm 1$ for $l=2$ as 
$|yz \rangle = \frac{\rm i}{\sqrt{2}} (|1 \rangle + |-1\rangle)$ 
and $|xz \rangle =- \frac{1}{\sqrt{2}} (|1 \rangle - |-1\rangle)$, respectively,   
the SOC $\lambda \vec l \cdot \vec s$ in the $\{yz,xz\}$ subset is written as 
\begin{eqnarray}
H_{\rm soc} =
\frac{\lambda}{2}  \sum_{\rm{\bf r} \, \sigma} 
\Bigl( {\rm i} \sigma_{\sigma \sigma}^z  c_{\rm{\bf r} a \sigma}^\dag c_{\rm{\bf r} b \sigma} + {\rm H.c.} \Bigr) , 
\label{eq:soc}
\end{eqnarray}
where $\hat \sigma^z$ is the $z$ component of the Pauli matrices. 

As shown in Supplementary Note 1, 
an effective model for the $\{xy,x^2-y^2\}$ doublet has the same form as the above $H_{\rm t} + H_{\rm soc}$.
By symmetry, there is no hopping matrix between the $\{yz,xz\}$ doublet and the other orbitals $xy$, $x^2-y^2$, and $3z^2-r^2$, 
but the $\{xy,x^2-y^2\}$ doublet and the $3z^2-r^2$ singlet could be hybridized. 
As discussed briefly later, the degeneracy in the $\{yz,xz\}$ doublet and in the $\{xy,x^2-y^2\}$ doublet could be lifted by a crystal field. 
Such band splitting is also induced by the difference between $t_\updelta$ and $t_\uppi$. 
Furthermore, all $d$ orbitals could in principle be mixed by the SOC. 
Including these complexities is possible but depends on the material and they usually induce smaller perturbations, therefore, here they are
left for future analyses.

\begin{figure}
\begin{center}
\includegraphics[width=0.8\columnwidth, clip]{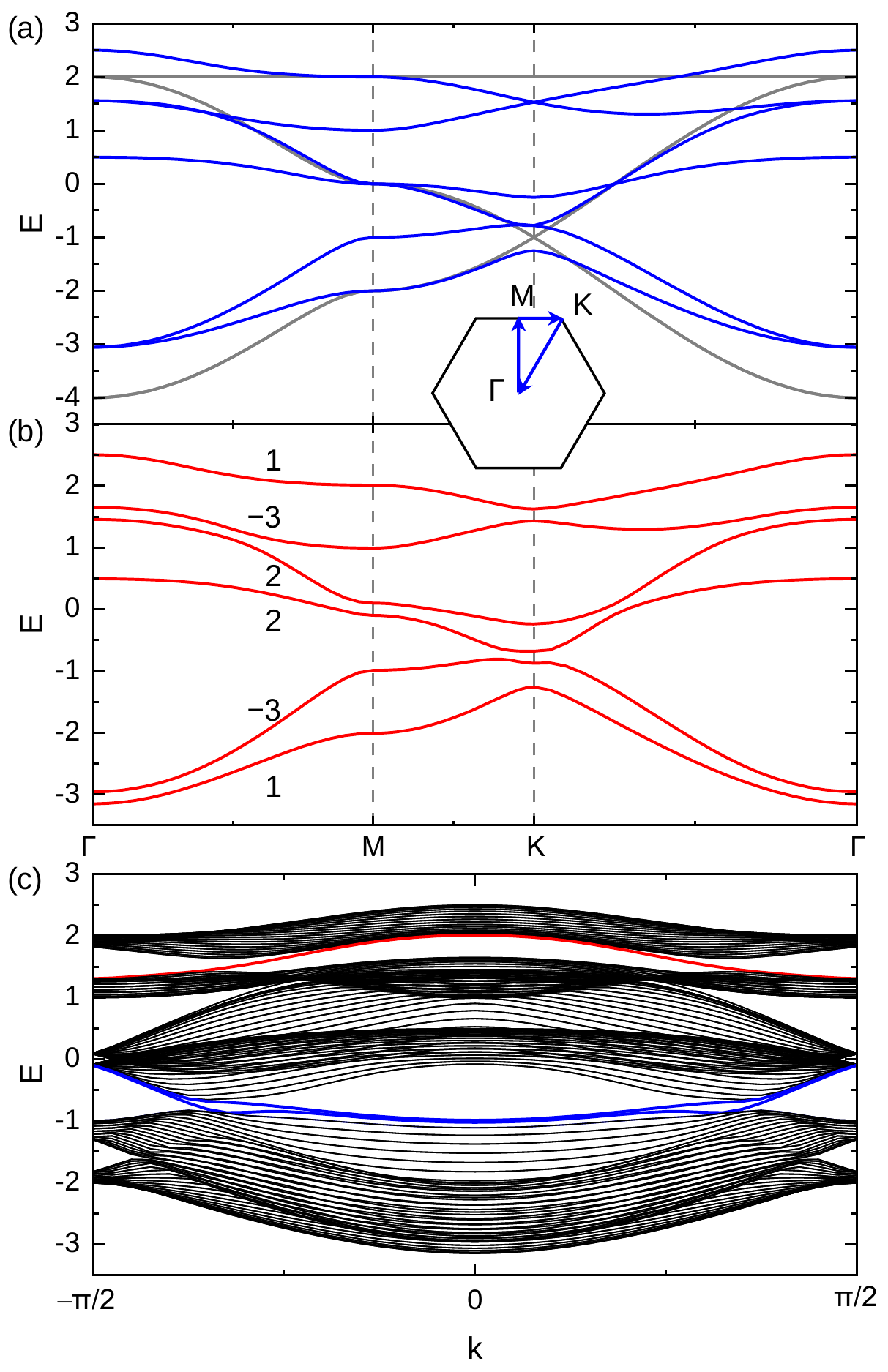}
\caption{Dispersion relations of the non-interacting model. 
Bulk dispersion relations 
without the spin-orbit coupling (SOC) (a) and with the SOC $\lambda=0.2$ (b).
In both cases, energy $E$ is scaled by the $\uppi$-bond hopping integral $t_\uppi$. 
Gray lines in (a) indicate the dispersion with the $\updelta$-bond hopping integral $t_\updelta=1$, 
which realizes the ideal dispersion in a kagome lattice. 
Blue lines in (a) and red lines in (b) are dispersions with $t_\updelta=0.5$. 
The inset shows the first Brillouin zone with high-symmetry lines used in panels (a) and (b). 
The Chern number ${\cal C}_n$ for the spin up component of each band is also shown in (b).
(c) Dispersion relations with $t_\updelta=0.5$ and $\lambda=0.2$ in the ribbon geometry, 
which is periodic along the ${\rm \bf a}_1$ direction and contains 20 unit cells along the perpendicular direction. 
Gapless edge modes are indicated by red and blue lines. 
}
\label{fig:singleparticle}
\end{center}
\end{figure}

{\flushleft \bf Non-interacting band topology.} 
By diagonalizing the single-particle Hamiltonian $H_{\rm t} + H_{\rm soc}$, one obtains dispersion relations as shown in Fig.~\ref{fig:singleparticle}. 
In the simplest case, where the hopping matrix $t_{\rm{\bf r} \, \rm{\bf r}'}^{\alpha \beta}$ does not distinguish $t_\updelta$ and $t_\uppi$ and the SOC is absent, 
the dispersion relation is identical to the one for the single-band tight-binding model, consisting of flat bands and graphene-like bands
as shown by gray lines in Fig.~\ref{fig:singleparticle} (a). 
Note that each band is fourfold degenerate because of two orbitals and two spins per site. 
Including SOC does not change the dispersion curve but simply shifts $\vec l \cdot \vec s=\pm1/2$ bands 
(see Supplementary Note 1). 

Including orbital dependence as $t_\updelta \ne t_\uppi$ without SOC instead splits the fourfold degeneracy except 
for two points at the $\Gamma$ point and two points at the $\rm K$ point. 
Quite intriguingly, Dirac dispersions emerge from the topmost flat bands as shown as blue lines in Fig.~\ref{fig:singleparticle} (a) 
(see Supplementary Note 1 for more discussion). 
Turning on the SOC further splits such fourfold degeneracy, leading to nontrivial band topology. 
In this particular example, the spin component along the $z$ axis is conserved giving unique characteristics to this case.
As shown in Fig.~\ref{fig:singleparticle} (b), the spin up component of each band is characterized by a nonzero Chern number ${\cal C}_n$. 
%
Because of the time-reversal symmetry, spin down bands have opposite Chern numbers. 
The topological property is also confirmed by gapless modes in the dispersion relation with the ribbon geometry, 
as shown in Fig.~\ref{fig:singleparticle} (c). 
Here, there appear one (two) pair of gapless modes between the highest and the second highest (between the second lowest and the third lowest) bands, 
shown as red (blue) curves, corresponding to the sum of Chern numbers below the gap, $-1 (-2)$. 
The other edge states are invisible because of the overlap with the bulk continuum.

A multi-orbital kagome model thus naturally shows quasi flat bands with nontrivial topology. 
However, close inspection revealed that, with $t_\updelta=0.5$ and $\lambda=0.2$, 
the minimum of the highest band at the $\rm K$ point is slightly lower than the maximum of the second highest band at the $\Gamma$ point. 
Thus, instead of a TI, a topological semimetal is realized 
when the Fermi level is located between the highest band and the second highest band.  
In fact, there are ways to make the gap positive. 
Here, we consider second-neighbor hopping matrices $\hat t^{(2)}_{\rm{\bf r} \, \rm{\bf r}'}$. 
As explained in Supplementary Note 1, these are also parametrized by 
$\uppi$-bonding $(dd\uppi)$ and $\updelta$-bonding $(dd\updelta)$, $t_\uppi^{(2)}$ and $t_\updelta^{(2)}$, respectively. 
For simplicity, we fix the ratio between $t_\uppi$ and $t_\uppi^{(2)}$ and between $t_\updelta$ and $t_\updelta^{(2)}$ as
$t_\uppi^{(2)}/t_\uppi=t_\updelta^{(2)}/t_\updelta = r_2$, and 
analyze the sign and magnitude of the band gap $\Delta_{\rm gap}$ between the highest band and the second highest band,
as well as the flatness of the highest band defined by $\Delta \varepsilon \equiv \varepsilon_{1, {\rm max}} - \varepsilon_{1, {\rm min}}$. 

\begin{figure}
\begin{center}
\includegraphics[width=0.8\columnwidth, clip]{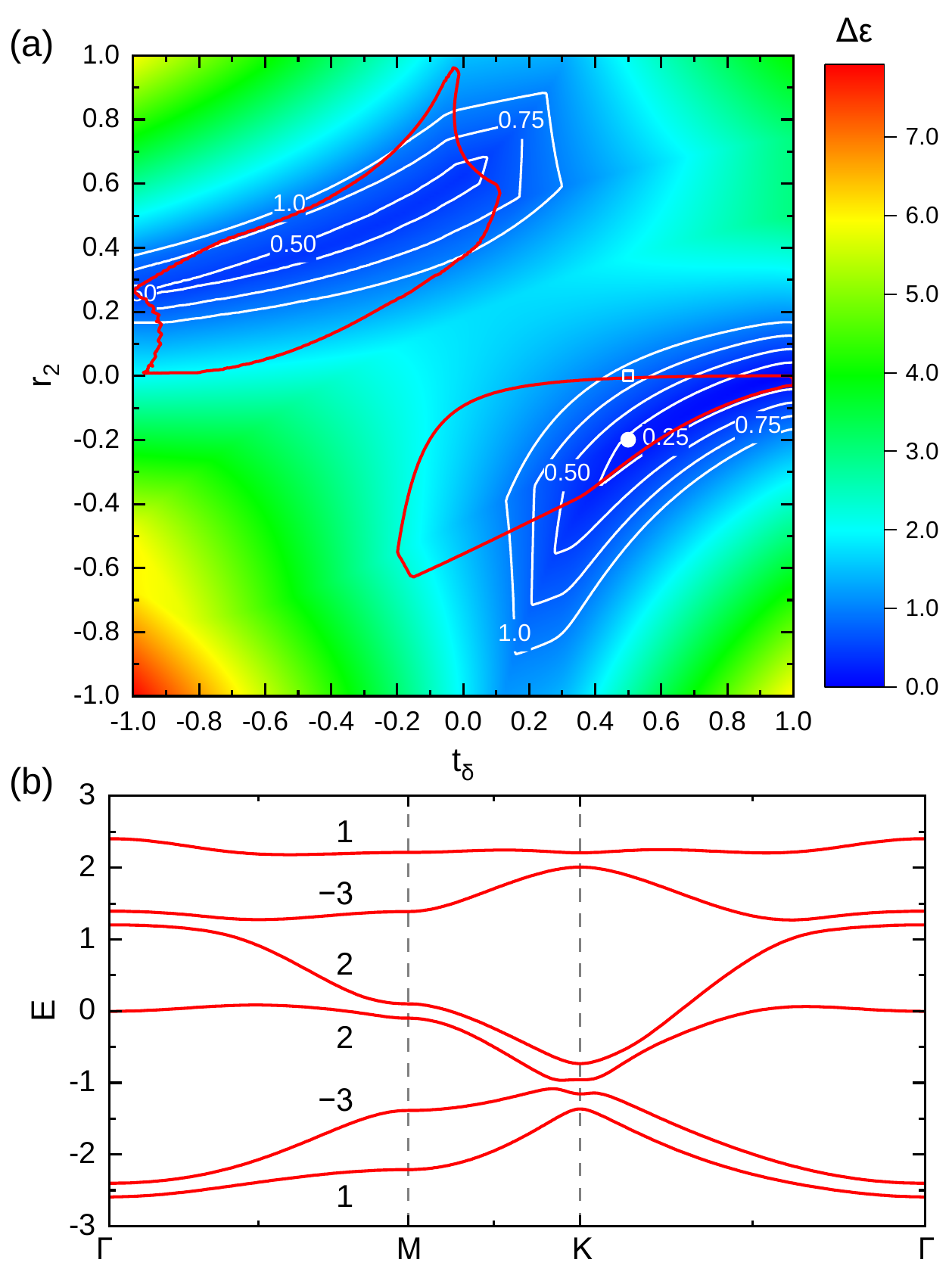}
\caption{Control of the band flatness. 
(a) Color map of the flatness $\Delta \varepsilon$ of the highest band as a function of $t_\updelta$ and 
the ratio between the nearest-neighbor and the second-neighbor hopping $r_2$ with $\lambda =0.2$. 
Open square (filled circle) locates $t_\updelta=0.5$ with $r_2=0$ ($-0.2$). 
Red closed loops show the areas where the band gap is positive $\Delta_{\rm gap}>0$. 
(b) Bulk band structure with $t_\updelta=0.5$, $r_2=-0.2$, and $\lambda=0.2$.
Band-dependent Chern number is also shown.}
\label{fig:diagram_r2}
\end{center}
\end{figure}

Figure~\ref{fig:diagram_r2} (a) plots $\Delta \varepsilon$ as a function of $t_\updelta$ and $r_2$ with $\lambda=0.2$. 
As mentioned previously, the perfectly flat band with $\Delta \varepsilon =0$ is realized at $t_\updelta=1$ and $r_2=0$, 
but band gap $\Delta_{\rm gap}$ is zero. 
The flatness is immediately modified by reducing $t_\updelta$ from 1.  
As indicated by an open square in the plot, $t_\updelta=0.5$ and $r_2=0$ gives 
$\Delta \varepsilon \sim 0.88$ and negative band gap $\Delta_{\rm gap} \sim -0.027$. 
Nonzero $r_2$ controls the relative energy between the zone center and the zone boundary. 
In particular, negative $r_2$ pushes up the energy at the $\rm K$ point, hereby the flatness is recovered. 
Naturally, the flatness and the positive gap are correlated as indicated by red loops in the second and forth quadrants 
because the separation between the highest band and the second highest band is fixed by the SOC strength. 
As indicated by a filled circle, $t_\updelta=0.5$ and $r_2=-0.2$ gives  
$\Delta \varepsilon \sim 0.22$ and positive band gap $\Delta_{\rm gap} \sim 0.17$. 
Corresponding dispersion relation is shown in Fig.~\ref{fig:diagram_r2} (b). 
The Chern numbers remain unchanged by this $r_2$. 

{\flushleft \bf Many-body effects.} 
%
Having established the topological properties at the single-particle level, 
we turn our attention to many-body effects focusing on the highest-energy flat band. 
A unique property of the current model is that the topmost quasi flat band has Chern number $|{\cal C}| =1$. 
Thus, a large spin polarization can be induced by many-body interactions \cite{Stoner1938} or by a small magnetic field. 
Further intriguing possibilities are FCI states 
when a topological flat band has a fractional filling and the insulating gap is induced by correlation effects 
\cite{Tang2011,Sun2011,Neupert2011,Sheng2011,Wang2011,Regnault2011,Xiao2011,Venderbos2012}. 
We examine such a possibility in our kagome model. 
Assuming the spin polarization in the highest band, 
we introduce local and nearest-neighbor Coulomb repulsive interactions as 
$H_{\rm U}=U \sum_{\rm{\bf r}} n_{\rm{\bf r} a \uparrow} n_{\rm{\bf r} b \uparrow} 
+ V \sum_{\langle \rm{\bf r} \rm{\bf r}' \rangle} 
\sum_{\alpha \beta} n_{\rm{\bf r} \alpha \uparrow} n_{\rm{\bf r}' \beta \uparrow}$, where 
$n_{\rm{\bf r} \alpha \sigma}=c_{\rm{\bf r} \alpha \sigma}^\dag c_{\rm{\bf r} \alpha \sigma}$. 
Here $U$ is the effective Coulomb interaction given by $U = U'-J$ 
with the interorbital Coulomb repulsion $U'$ and the interorbital exchange interaction $J$. 
These interactions are then projected onto the highest band, leading to the effective Hamiltonian $H_{\rm eff} = H_{\rm t}+H_{\rm soc}+H_{\rm U}$.

Note that the $S_z$ conservation is not essential to realize FCI. 
For our case and most of others, including complexities which break $S_z$ conservation does not destroy FCI 
as long as the flat band has the nontrivial topology and is well separated from other bands, 
justifying projecting interaction terms onto the flat band and allowing for an accurate Lanczos calculation.
While computational cost would be expensive, direct calculations of multiband models with $S_z$-non-conserving terms 
would show FCI if the appropriate condition is fulfilled, but this possibility has not been fully explored yet. 

\begin{figure}
\begin{center}
\includegraphics[width=0.8\columnwidth, clip]{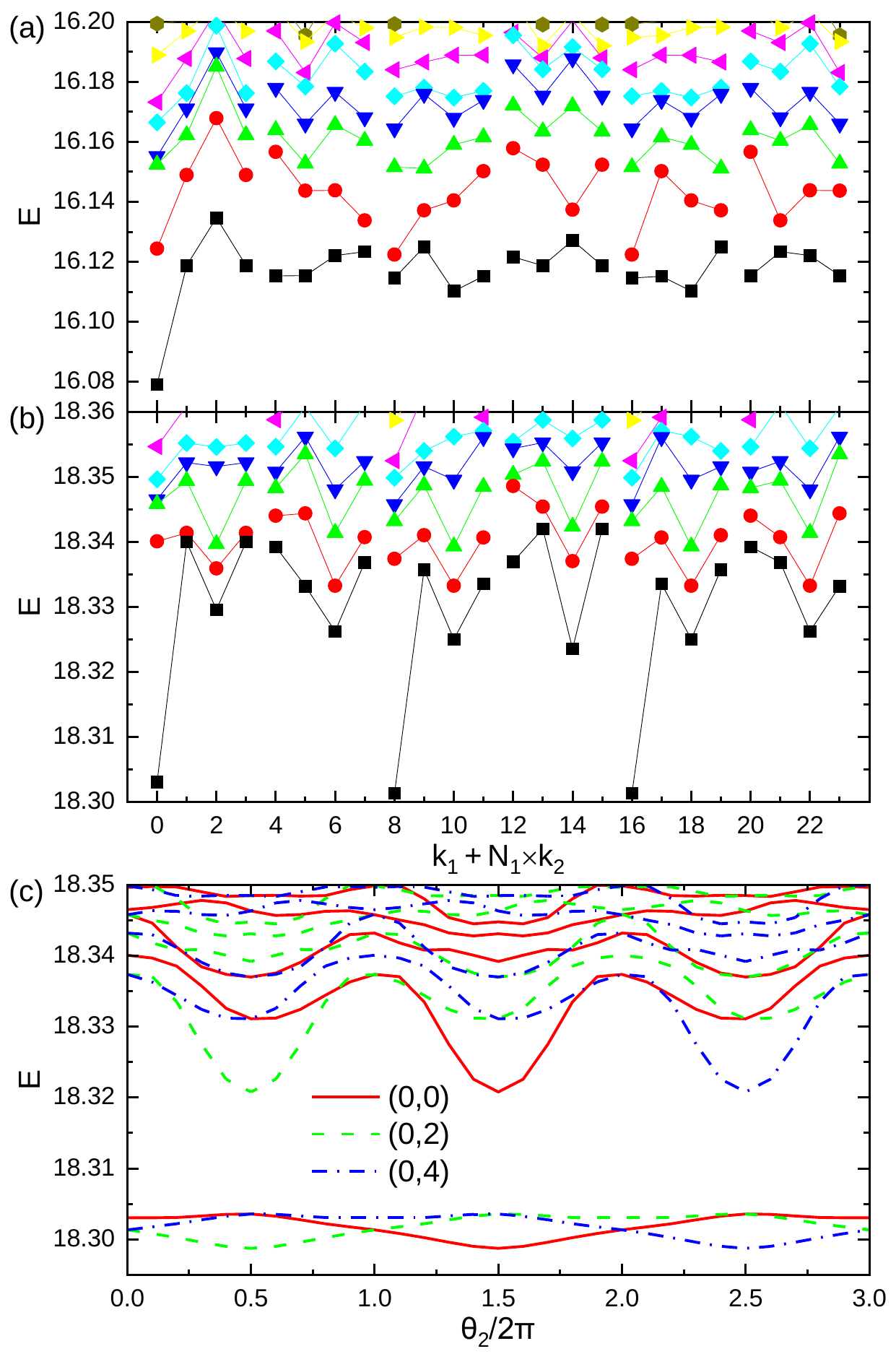}
\caption{Emergence of a $\nu=1/3$ fractional Chern insulating state. 
Low-energy spectra of an interacting model with $t_\updelta=0.5$ with $r_2=0$ (a) and $r_2=-0.2$ (b). 
Other parameter values are $\lambda=0.2$, $U=2$ (local Coulomb interaction), and $V=1$ (nearest-neighbor Coulomb interaction). 
Ground state energy is indicated by black squares, and excited state energies are indicated by different symbols.  
(c) Spectral flow of the ground state manifold upon flux insertion with $t_\updelta=0.5$ with $r_2=-0.2$. 
Red solid lines, green dashed lines, and blue dash-dotted lines are for sector $(k_1,k_2)=(0,0)$, $(0,2)$, and $(0,4)$, respectively. 
}
\label{fig:fqh}
\end{center}
\end{figure}

The effective Hamiltonian $H_{\rm eff}$ is diagonalized in momentum space. 
For this purpose, we discretize the momentum space into $N_1 \times N_2$ patches and express the Hamiltonian in the occupation basis, 
i.e., the Hilbert space is built up by $|\varphi_l \rangle=\prod_{{\rm \bf k}\in l} \psi^\dag_{1 {\rm \bf k}} |0\rangle$, 
where $\psi_{1 {\rm \bf k}}$ is the single-particle wave function for the highest flat band at momentum $\rm \bf k$, 
and the combination of $\rm \bf k$ is specified by $l$. 
Due to the translational symmetry and the momentum conservation of many-body interaction terms, 
$H_{\rm eff}$ is subdiagonalized according to  the total momentum 
${\rm \bf k}_{\rm tot}= \sum_{\rm{\bf k} \in l} {\rm \bf k}$ 
modulo 
$\rm{\bf b}_1$ and $\rm{\bf b}_2$, with $\rm{\bf b}_{1,2}$ being two reciprocal lattice vectors. 
In this study, we take $N_1=4$ and $N_2=6$ and consider $\nu=1/3$ filling, that is, the number electrons in the highest flat band is
 $N_{\rm e}=8$. 
Momentum sector will be specified using integer index $(k_1,k_2)$ corresponding to the total momentum 
${\rm \bf k}_{\rm tot}= {\rm \bf b}_1 k_1/ N_1 +  {\rm \bf b}_2 k_2/ N_2$. 

Figures~\ref{fig:fqh} (a) and (b) show the low-energy spectra of the interacting model with $t_\updelta=0.5$ and $r_2=0$ 
and $t_\updelta=0.5$ and $r_2=-0.2$, respectively, 
with $U=2$ and $V=1$ as a function of total momentum.  
In (a), the energy spectrum has a unique ground state at total momentum $(k_1,k_2)=(0,0)$ 
(note that this is to show the competition between the wide band width and the correlation effects on the highest band). 
When $r_2$ is introduced as $-0.2$, the highest band becomes flatter, leading to a drastic change in the energy spectrum. 
There appear three energy minima at $(k_1,k_2)=(0,0)$, $(0,2)$, and $(0,4)$, 
forming a threefold degenerate ground state manifold (GSM), 
which is separated from the other states by an energy gap $\sim 0.03$. 
As shown in Fig.~\ref{fig:fqh} (c), the three sectors evolve with each other by inserting magnetic fluxes without having overlap with higher energy states (energy separation is slightly reduced to $\sim 0.02$). 
These results strongly suggest a $\nu=1/3$ FCI state.

To confirm that this threefold degenerate ground state really represents a FCI state instead of trivial states such as charge density waves, 
we compute Chern numbers ${\cal C}_{(k_1,k_2)}$ by introducing twisted boundary conditions \cite{Thouless1982,Niu1985}. 
Here, we discretize the boundary phase unit cell into $20 \times 20$ meshes, and numerically evaluate the Berry curvature $F_{(k_1,k_2)}(\theta_1,\theta_2)$ 
as detailed in the Methods section 
as well as in Supplementary Note 2. 
Figure~\ref{fig:manybodyberry} shows $F_{(k_1,k_2)}(\theta_1,\theta_2)$ in a discretized grid $(n_1,n_2)$ for the ground state manifold with $t_\updelta=0.5$, $r_2=-0.2$, $\lambda=0.2$ with $U=2$ and $V=1$. 
Along the $n_1$ direction, these plots are periodic. 
Along the $n_2$ direction, plot (a) is continuously connected to plot (b), plot (b) is connected to plot (c), and plot (c) is connected back to plot (a). 
This also confirms the threefold ground state manifold, where inserting one flux quantum along the ${\rm \bf b}_2$ direction shifts the sector $(k_1,k_2)=(0,0)$ to $(0,2)$, $(0,2)$ to $(0,4)$, and $(0,4)$ to $(0,0)$. 
%
By adding up the discretized values of $F_{(k_1,k_2)}(\theta_1,\theta_2)$,
we obtain ${\cal C}_{(0,0)} = 0.331489$, ${\cal C}_{(0,2)} =0.330318$, ${\cal C}_{(0,4)} =0.338193$, 
and the sum of the three Chern numbers is exactly 1 within the numerical accuracy.
The slight deviation from the ideal value ${\cal C} = 1/3$ is ascribed to finite-size effects. 
This proves the existence of a $\nu = 1/3$ FCI phase with a quantized fractional Hall response $\sigma_{\rm H}=\frac 1 3 e^2/h$, 
where $e$ is the electron charge and $h$ is the Planck constant. 
In our numerical analyses, 
we did not find a ground state with the threefold degeneracy and Chern number zero, thus excluding the charge density wave states. 
This is probably because the quantum effects make such states unstable, as discussed in \cite{Yoshioka1983}. 

\begin{figure*}
\begin{center}
\includegraphics[width=1.7\columnwidth, clip]{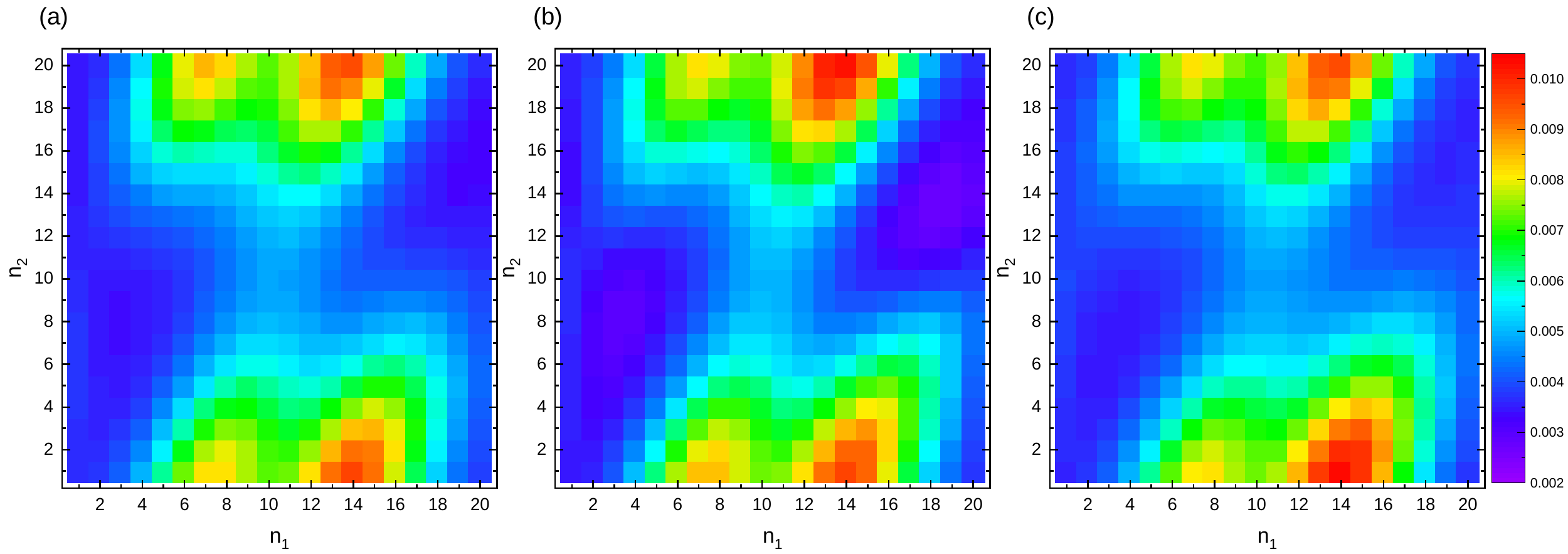}
\caption{Many-body Berry curvature as a function of discretized boundary phases. 
(a) Sector $(k_1,k_2)=(0,0)$, (b) sector $(0,2)$, and (c) sector $(0,4)$.
Parameter values are $t_\updelta=0.5$, $r_2=-0.2$, $\lambda =0.2$ with $U=2$ and $V=1$.  
}
\label{fig:manybodyberry}
\end{center}
\end{figure*}

\section*{Discussion}
%
In this work, we have considered an itinerant electron model on a kagome lattice with twofold degenerate orbitals per site. 
However, each site has $C_2$ rotational symmetry, rather than $C_3$ or $C_4$. 
Thus, the degeneracy between the two orbitals ($yz$ and $xz$) can be lifted. 
In our tight-binding model, a difference in the hopping amplitude between $t_\pi$ and $t_\delta$ in fact 
lifts such degeneracy, leading to the splitting of the band structure. 
Thus, adding local crystal field splitting, which respects the underlying lattice symmetry, would not fully destroy the topological property found in this work, 
while the position of topological or flat bands would be modified depending on model parameters. 
As a number of kagome materials have already displayed a nontrivial band topology 
\cite{Alled2012,Yin2019,Jiao2019,Liu2019,Ye2018,Lin2018,Kan2020,Nakatsuji2015,Kuroda2017,Nayak2016,Ortiz2020,Wu2021,Feng2021,Denner2021}, 
reducing the thickness of such materials down to a few unit cells, or growing thin films of such materials 
and tuning the Fermi level to a topological flat band by chemical substitution or gating, 
might be a promising route to observe the phenomena predicted here. 
The sign and the magnitude of the parameter $r_2$ could depend on details of the material, such as the species of ligand ions, and 
might be further controlled by compressive or tensile strain. 
First principles calculations would help to construct realistic material-dependent models \cite{Liu2013,Zhou2014,Yamada2016}. 
It is anticipated that the separation between $\{yz,xz\}$, $\{xy,x^2-y^2\}$, or $3z^3-r^2$ subsets will be enhanced by reducing the film thickness 
compared with that in the bulk so that one can focus on one of the subsets only.  
In addition to a kagome lattice, topological flat bands appear in dice and Lieb lattices \cite{Wang2011b,Soni2020,Soni2021}. 
Study of FCI states in such lattice geometries and material search is another important direction. 

To summarize, 
we have demonstrated the close interplay between the spatial frustration and the orbital degree of freedom in a kagome lattice. 
With the relativistic spin-orbit coupling, such an interplay not only affects the band dispersion, but also induces nontrivial topology. 
Specifically, we showed that the original flat bands in a kagome lattice become dispersive and topologically nontrivial. 
When such topological bands are fractionally occupied by electrons, many-body interactions drive 
further intriguing phenomena, i.e., fractional Chern insulating states. 
Our work may bridge the gap between idealized theoretical studies and real materials. 

\section*{Methods}

{\flushleft \bf Non-interacting $\{yz,xz\}$ model.} 
Here we deduce the hopping matrices of the $\{yz,xz\}$ model in the Slater-Koster approximation. 

For nearest-neighbor bonds, in addition to the diagonal matrix $\hat t_{\bf 1 \, 2}$ presented in the main text, we have 
\begin{eqnarray}
&\hat t_{\bf 1 \, 3} = \frac{1}{4} 
\left[
\begin{matrix}
3 t_\uppi + t_\updelta & \sqrt{3} (t_\pi - t_\updelta) \\
\sqrt{3} (t_\uppi - t_\updelta) & t_\uppi + 3 t_\updelta
\end{matrix}
\right],& \nonumber \\ 
&\hat t_{\bf 2 \, 3} = \frac{1}{4} 
\left[
\begin{matrix}
3 t_\uppi + t_\updelta &  - \sqrt{3} (t_\uppi - t_\updelta) \\
- \sqrt{3} (t_\uppi - t_\updelta) & t_\uppi + 3 t_\updelta
\end{matrix}
\right].& 
\end{eqnarray}

\begin{figure}
\begin{center}
\includegraphics[width=0.6\columnwidth, clip]{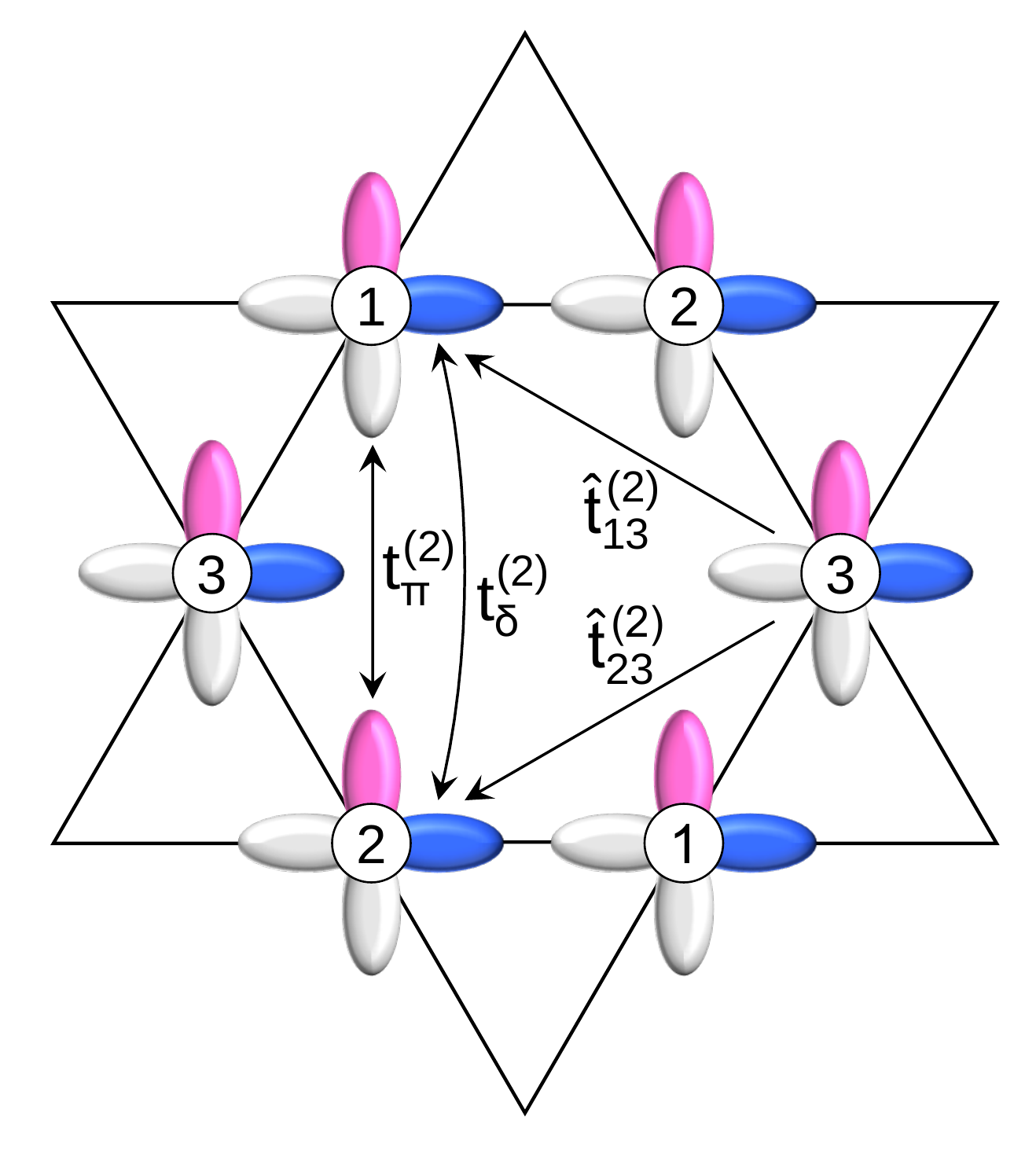}
\caption{Second neighbor hopping matrices. 
$yz (xz)$ orbitals between site 1 and site 2 are hybridized via diagonal hopping $t^{(2)}_{\uppi (\updelta)}$, i.e., $\uppi (\updelta)$ bonding. 
Other hopping integrals between site 1 and site 3 and between site 2 and site 3 are given by 
$\hat t_{\bf 1 \, 3}^{(2)}$ and $\hat t_{\bf 2 \, 3}^{(2)}$, respectively, 
obtained via the Slater rule\cite{Slater1954}. 
}
\label{fig:secondneighbor}
\end{center}
\end{figure}

Similarly, second neighbor hopping matrices can be written as 
\begin{eqnarray}
&\hat t_{\bf 1 \, 2}^{(2)} = 
\left[
\begin{matrix}
t_\uppi^{(2)}  & 0 \\
0 &  t_\updelta^{(2)}
\end{matrix}
\right],& \nonumber \\ 
&\hat t_{\bf 1 \, 3}^{(2)} = \frac{1}{4} 
\left[
\begin{matrix}
t_\uppi^{(2)} + 3 t_\updelta^{(2)} & -\sqrt{3} (t_\uppi^{(2)} - t_\updelta^{(2)}) \\
-\sqrt{3} (t_\uppi^{(2)} - t_\updelta^{(2)}) & 3 t_\uppi^{(2)} + t_\updelta^{(2)}
\end{matrix}
\right],& \nonumber \\ 
&\hat t_{\bf 2 \, 3}^{(2)} = \frac{1}{4} 
\left[
\begin{matrix}
t_\uppi^{(2)} + 3 t_\updelta^{(2)} & \sqrt{3} (t_\uppi^{(2)} - t_\updelta^{(2)}) \\
\sqrt{3} (t_\uppi^{(2)} - t_\updelta^{(2)}) & 3 t_\uppi^{(2)} + t_\updelta^{(2)}
\end{matrix}
\right],& 
\end{eqnarray}
where subscript $(2)$ is introduced to highlight the difference from the nearest-neighbor bonds. 
These are schematically shown in Fig.~\ref{fig:secondneighbor}. 
$t_\uppi^{(2)}$ and $t_\updelta^{(2)}$ correspond to $(dd\uppi)$ and $(dd\updelta)$, respectively, by Slater and Koster\cite{Slater1954}. 

{\flushleft \bf Non-interacting Berry curvature.}
The band-dependent Berry curvature of non-interacting electrons is given as a function of momentum $\rm \bf k$ as 
\begin{equation}
\Omega_{n {\rm \bf k}}={\rm i} 
\sum_{m (\ne n)} \frac{\langle n| \hat v_{x {\rm \bf k}}|m \rangle \langle m| \hat v_{y {\rm \bf k}} |n \rangle 
- (\hat v_{x {\rm \bf k}} \leftrightarrow \hat v_{y {\rm \bf k}}) 
}{(\varepsilon_{m {\rm \bf k}} - \varepsilon_{n {\rm \bf k}})^2}, 
\end{equation}
where, using the Hamiltonian matrix in momentum space $\hat H_{\rm \bf k}$,  
$\hat v_{\eta {\rm \bf k}}$ is given by $\hat v_{\eta {\rm \bf k}} = \partial \hat H_{\rm \bf k}/\partial k_\eta$. 
With this Berry curvature, the band dependent Chern number ${\cal C}_n$ is given by 
\begin{equation}
{\cal C}_n =\frac{1}{2 \uppi} \int_{\rm BZ} d^2k \, \Omega_{n {\rm \bf k}}, 
\end{equation}
where the momentum integral is taken in the first Brillouin zone. 

{\flushleft \bf Many-body Chern number.}
The many-body Chern number is computed by introducing a twist boundary condition to a single-particle wave function as  
$\psi ({\rm{\bf r}} + N_j {\rm{\bf a}}_j) = e^{{\rm i} \theta_j}\psi (\rm{\bf r})$, 
where 
$N_{j=1,2}$ are the numbers of unit cells along lattice translation vectors ${\rm \bf a}_{j=1,2}$, 
with phase factors $\theta_{j=1,2}$. 
This corresponds to inserting magnetic fluxes. 
When one flux quantum is inserted, $\theta_j$ changes from $0$ to $2\pi$ 
and discretized momentum $\rm \bf k$ moves from its original position to its neighbor along the  ${\rm \bf b}_j$ direction 
with the momentum shift given by $\Delta {\rm \bf k} = {\rm \bf b}_j /N_j$. 

Many-body Chern number of the ground state $(k_1,k_2)$ is computed via 
${\cal C}_{(k_1,k_2)} = \frac{1}{2 \uppi} \! \int_0^{2 \uppi} \! d \theta_1 \int_0^{2 \uppi} \! d \theta_2 F_{(k_1,k_2)} (\theta_1, \theta_2)$ \cite{Niu1985}
where $F(\theta_1, \theta_2)$ is the Berry curvature given by
\begin{equation}
F_{(k_1,k_2)}(\theta_1, \theta_2)= {\rm Im} \biggl\{
\bigg\langle \frac{\partial \Phi_{(k_1,k_2)}}{\partial \theta_2} \bigg| \frac{\partial \Phi_{(k_1,k_2)}}{\partial \theta_1} \bigg\rangle 
- (\theta_1 \leftrightarrow \theta_2)
\biggr\}. 
\label{eq:berry}
\end{equation}
Here, $| \Phi_{(k_1,k_2)} \rangle$ is the many-body wave function constructed using single-particle wave functions with a twist boundary condition 
$\psi ({\rm{\bf r}})$ after the Fourier transformation to momentum space. 
The momentum index $(k_1,k_2)$ will be omitted in the following discussion for simplicity. 

Partial derivative of a wave function with respect to $\theta_j$ is approximated by a finite difference 
as $|\partial \Phi/\partial \theta \rangle \approx \frac{1}{|\Delta {\boldsymbol \theta}|} 
[| \Phi ({\boldsymbol \theta} + \Delta {\boldsymbol \theta}) \rangle - | \Phi ({\boldsymbol \theta}) \rangle]$. 
Here, the vector notation is used for ${\boldsymbol \theta} = (\theta_1,\theta_2)$, and $\Delta {\boldsymbol \theta} = (\Delta \theta_1,0)$ or $(0,\Delta \theta_2)$. 
Then, it is required to compute a product of two wave functions as 
$\langle \Phi ({\boldsymbol \theta})|\Phi ({\boldsymbol \theta}') \rangle$ with $\boldsymbol \theta \ne \boldsymbol \theta'$. 
Because we are using a multiorbital model projected onto the flat band, special care is needed, as detailed in Supplementary Note~2. 

\section*{Data availability}
The data that support the findings of this study are available from the corresponding
author upon reasonable request.

\section*{Code availability}
Codes used in this paper are available from the corresponding author upon reasonable request.

\section*{Acknowledgments}
The research of S.O., N.M., and E.D. was supported by the  U.S. Department of Energy, Office of Science, Basic Energy Sciences, Materials Sciences and Engineering Division. 
D.N.S was supported by the U.S. Department of Energy, Office of Basic Energy Sciences under Grant No. DE-FG02-06ER46305 for numerical studies of topological interacting systems.
S.O. thanks H. Miao and H. Li for discussions. 
This research used resources of the Compute and Data Environment for Science (CADES) at the Oak Ridge National Laboratory, 
which is supported by the Office of Science of the U.S. Department of Energy under Contract No. DE-AC05-00OR22725. 

Copyright  notice: This  manuscript  has  been  authored by UT-Battelle, LLC under Contract No. DE-AC05-00OR22725 with the U.S.  Department  of  Energy.   
The  United  States  Government  retains  and  the  publisher,  by  accepting  the  article  for  publication, 
acknowledges  that  the  United  States  Government  retains  a  non-exclusive, paid-up, irrevocable, world-wide license to publish or reproduce the published form of this manuscript, 
or allow others to do so, for United States Government purposes.  
The Department of Energy will provide public access to these results of federally sponsored  research  in  accordance  with  the  DOE  Public  Access  Plan 
(http://energy.gov/downloads/doe-public-access-plan)

\section*{Author contributions}

S.O. designed the research and carried out numerical calculations and wrote the manuscript with the input from all the authors. 
N.M. supported the construction of the model Hamiltonian. 
E.D. and D.N.S. supported many-body numerical calculations. 

\section*{Competing interests}
The authors declare that there are no competing interests.

\setcounter{MaxMatrixCols}{10}


\onecolumngrid

\newpage
\begin{center}
{\large \bf Supplementary Information: Topological flat bands in a  kagome lattice multiorbital system}\\
\vspace{1em}
Satoshi Okamoto,$^1$ Narayan Mohanta,$^1$ Elbio Dagotto,$^{1,2}$ and D. N. Sheng$^3$ \\
\vspace{0.5em}
{\small \it $^1$Materials Science and Technology Division, Oak Ridge National Laboratory, Oak Ridge, Tennessee 37831, USA}\\
{\small \it $^2$Department of Physics and Astronomy, The University of Tennessee, Knoxville, Tennessee 37996, USA}\\
{\small \it $^3$Department of Physics and Astronomy, California State University, Northridge, California 91330, USA}\\
\end{center}

\setcounter{page}{1}
\renewcommand{\thetable}{S\Roman{table}}
\renewcommand{\thefigure}{S\arabic{figure}}
\renewcommand{\thesection}{Supplementary Note \arabic{section}}
\renewcommand{\thesubsection}{\arabic{subsection}}
\renewcommand{\thesubsubsection}{\arabic{subsection}.\arabic{subsubsection}}
\renewcommand{\theequation}{S\arabic{equation}}

\setcounter{secnumdepth}{3}

\setcounter{equation}{0}
\setcounter{figure}{0}

\section{Non-interacting case}

\subsection{$\{yz,xz\}$ model}

\begin{figure}[h]
\begin{center}
\includegraphics[width=0.4\columnwidth, clip]{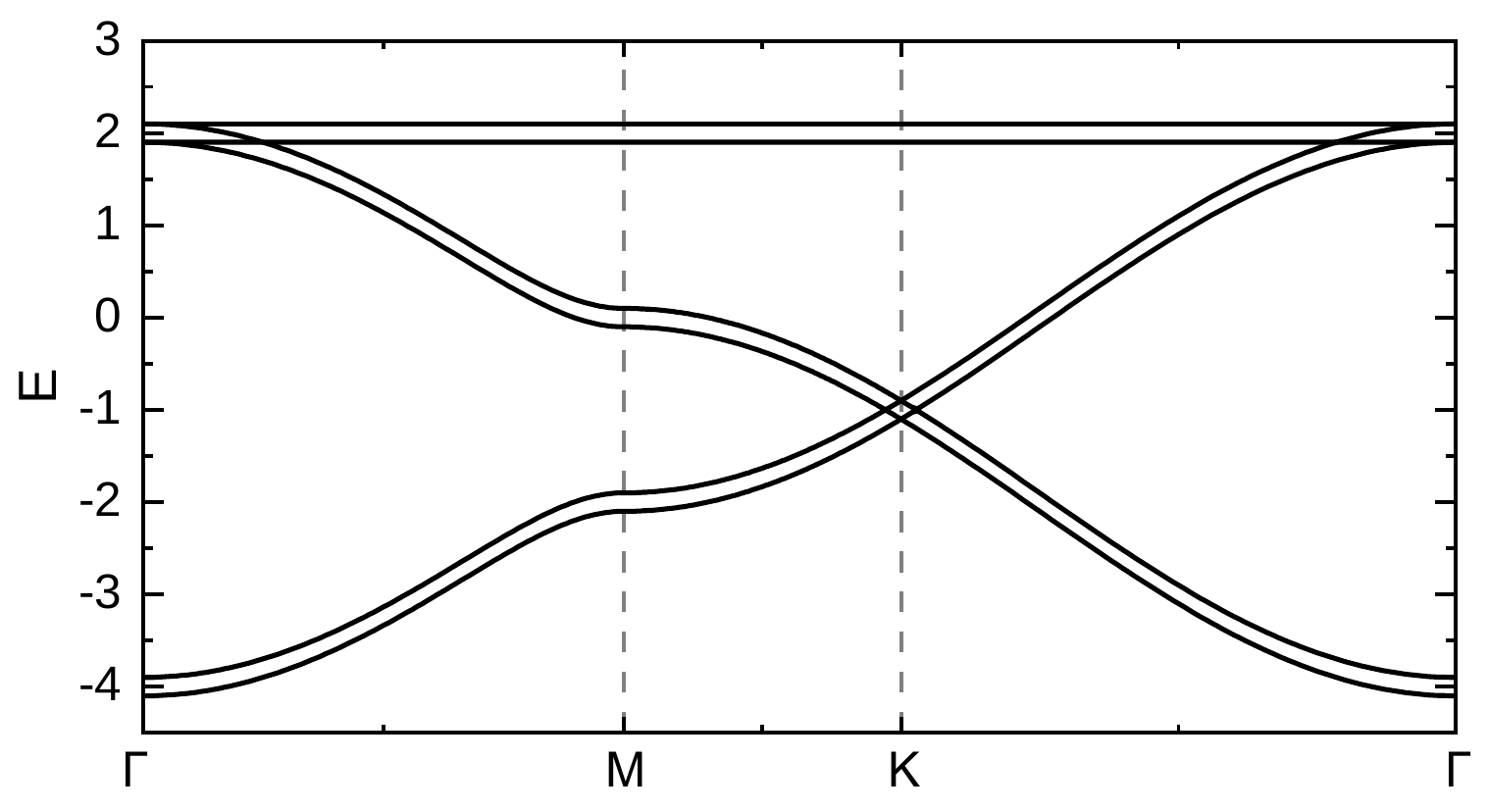}
\caption{Bulk dispersion relation of the non-interacting model. 
Energy is scaled by $\uppi$-bond hopping $t_\uppi=1$. 
Other parameters are $\updelta$-bond hopping $t_\updelta=1$, 
the ratio between nearest-neighbor and second-neighbor hopping $r_2=0$, and the spin-orbit coupling $\lambda=0.2$. 
}
\label{fig:bulkdispersion_r11r20}
\end{center}
\end{figure}

Bulk dispersion relation for the simplest case of the $\{yz,xz\}$ model with $t_\updelta=1$, $r_2=0$, and $\lambda=0.2$ 
is presented in Fig.~\ref{fig:bulkdispersion_r11r20}. 
With this set of parameters, the dispersion relation consists of two dispersions of ideal kagome system with nearest-neighbor hopping separated by $\lambda$ as each set of dispersions comes from 
$\vec l \cdot \vec s =1/2$ or $\vec l \cdot \vec s =-1/2$ states. 

\begin{figure}[h]
\begin{center}
\includegraphics[width=0.3\columnwidth, clip]{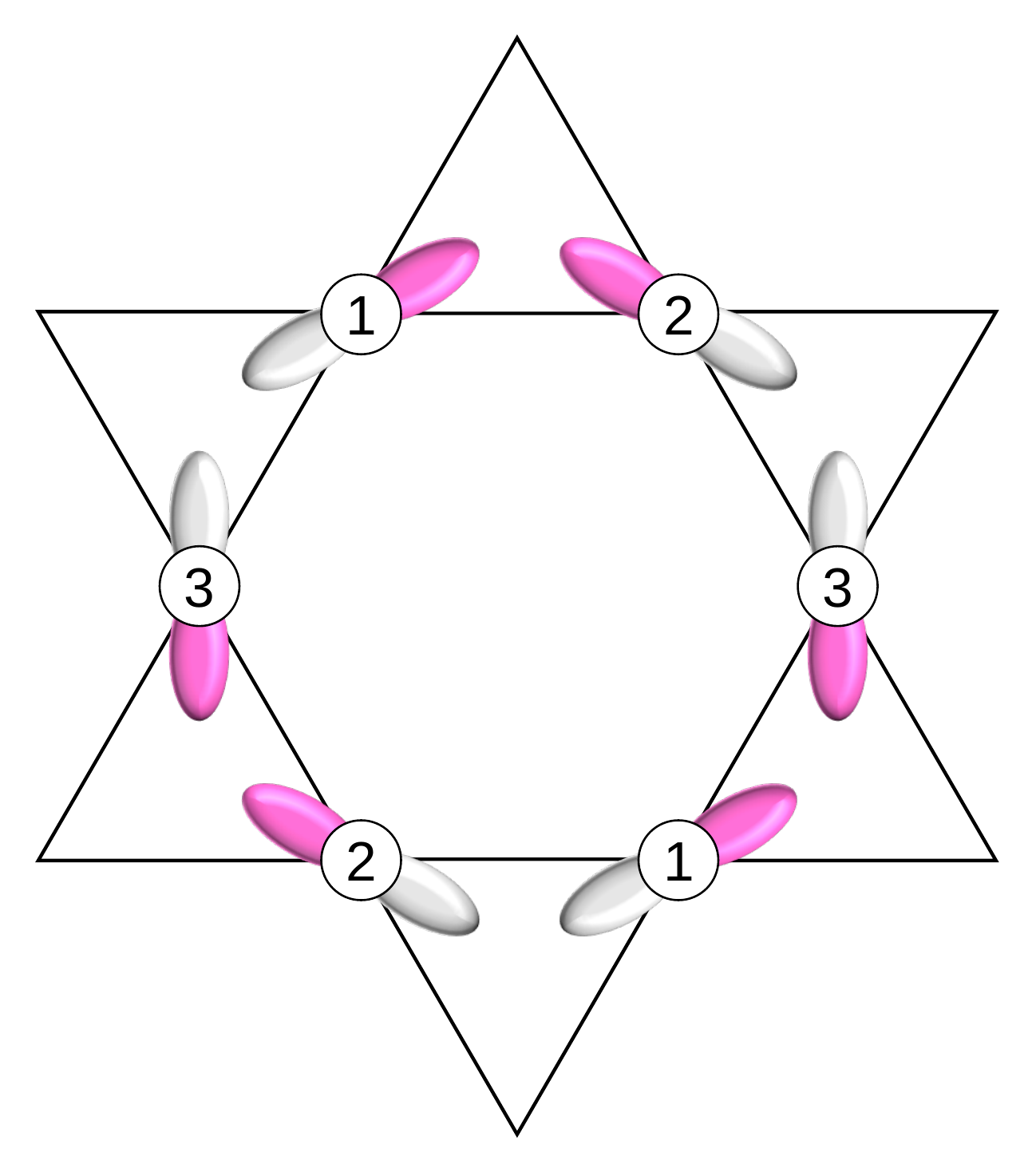}
\caption{Orbital configuration of the top flat band at the $\Gamma$ point.
Colored ellipsoids indicate regions where electron wave functions have positive sign at positive $z$ direction (perpendicular to the plane). 
$t_\updelta$ is nonzero, but the spin-orbit coupling is turned off $\lambda =0$. 
}
\label{fig:orbitalGamma}
\end{center}
\end{figure}

How Dirac dispersions emerge from the top flat bands by introducing $t_\updelta \ne 1$? 
While the non-zero momentum and the spin-orbit coupling induce complexity in the system, 
qualitatively the primary contribution to the top flat band is from a linear and symmetric combination of $yz$ and $xz$ orbitals. 
At the $\Gamma$ point and without the spin-orbit coupling, this is explicitly given by 
\begin{eqnarray}
\frac{1}{\sqrt{6}}
\bigl(
\cos \eta_1 |yz, 1 \rangle + \sin \eta_1 |xz, 1 \rangle+\cos \eta_2 |yz, 2 \rangle + \sin \eta_2 |xz, 2 \rangle+\cos \eta_3 |yz, 3 \rangle + \sin \eta_3 |xz, 3 \rangle
\bigr) 
\end{eqnarray}
with
$\eta_1=60^\circ$, $\eta_2=300^\circ$, $\eta_3=180^\circ$. 
Here, $|a, l \rangle$ stands for orbital $a$ at sublattice $l$. 
This is schematically shown in Fig.~\ref{fig:orbitalGamma}. 
The hybridization with other combinations of orbitals induces band dispersions and Dirac band crossing. 
Similar wave functions are obtained from density functional theory calculations on CoSn in \cite{Kang2020Sup}. 

\begin{figure}[h]
\begin{center}
\includegraphics[width=1\columnwidth, clip]{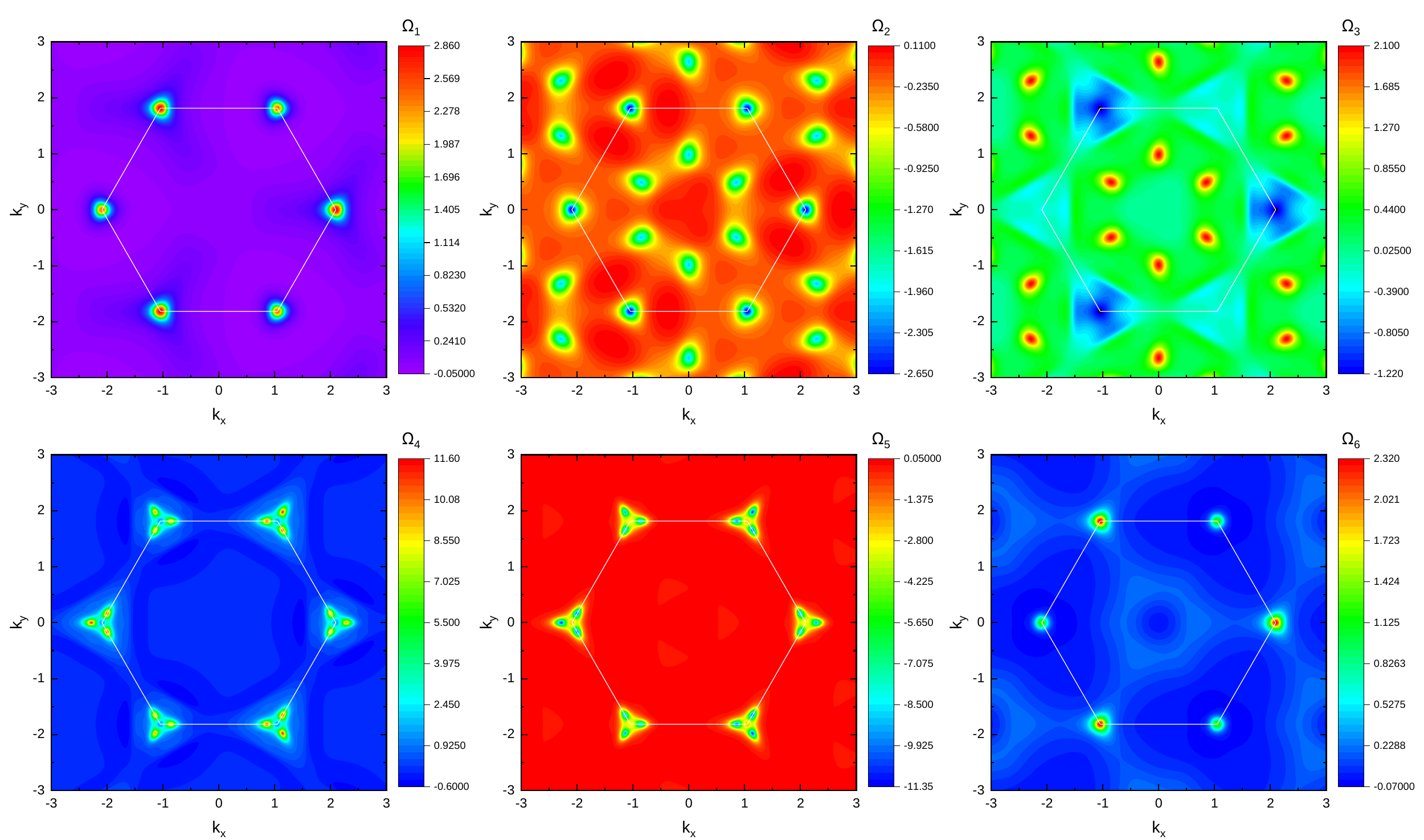}
\caption{Berry curvature $\Omega_i$ of non-interacting model as a function of two-dimensional momentum $(k_x,k_y)$. 
$\Omega_1$ is for the highest up band. 
Parameter values are $t_\updelta=0.5$ and $\lambda=0.2$ 
White lines indicate the first Brillouin zone. 
Berry curvature of down bands have negative value with the sign of momentum changed, 
$\Omega_i (k_x,k_y) \rightarrow -\Omega_i (-k_x,-k_y)$.
Here, Berry curvature is divided by $2 \uppi$, so that the momentum integral over the first Brillouin zone equals the Chern number. }
\label{fig:berry_td05}
\end{center}
\end{figure}

Figure~\ref{fig:berry_td05} shows the band-dependent Berry curvature for $t_\updelta=0.5$ and $r_2=0$ with $\lambda=0.2$, a parameter set used in Figs. 2 (b) and (c) in the main text. 
Nonzero $r_2$ mostly affects the Berry curvature of dispersive bands because it changes the band dispersion curves of these bands. 
On the other hand, the Berry curvature of the top flat band is qualitatively unchanged because of the following reasons: 
1. the top band is already flat without including $r_2$, and 
2. the Berry curvature peaks near the K point, where the separation from the second top band is minimum, 
but the separation is fixed by the SOC strength. 
The momentum integral of each plot gives band-dependent Chern number as shown in Fig. 2 (b).

\begin{figure}[h]
\begin{center}
\includegraphics[width=0.8\columnwidth, clip]{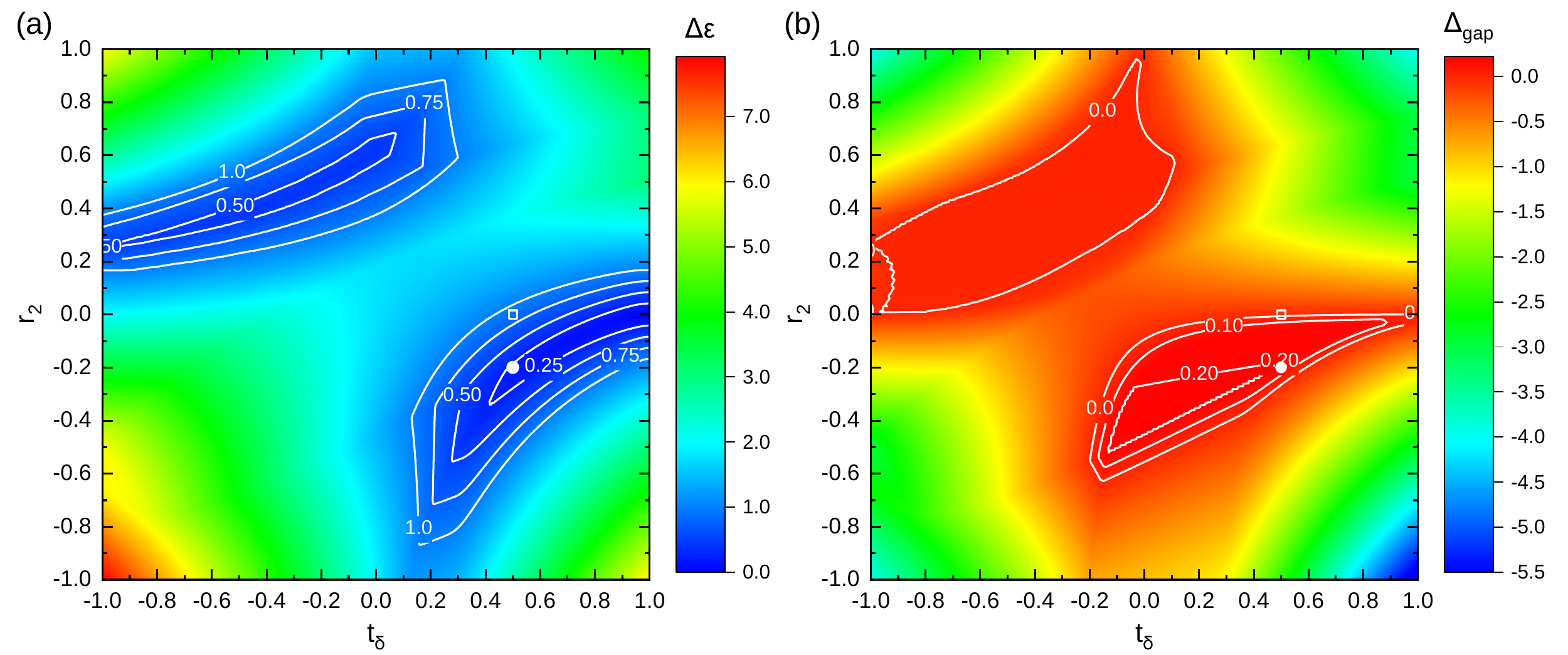}
\caption{Control of the band flatness. 
(a) Flatness $\Delta \varepsilon$ of the highest band and (b) gap amplitude $\Delta_{\rm gap}$ between the highest band and the second highest band as 
functions of $t_\updelta$ and $r_2$ with $\lambda=0.2$. 
}
\label{fig:plot_r2}
\end{center}
\end{figure}

To examine the role of second-neighbor hoppings, 
we compute the flatness of the highest band defined by $\Delta \varepsilon \equiv \varepsilon_{1, {\rm max}} - \varepsilon_{1, {\rm min}}$ 
and the band gap defined by $\Delta_{\rm gap} \equiv \varepsilon_{1, {\rm min}} - \varepsilon_{2, {\rm max}}$.
Results are summarized in Figure~\ref{fig:plot_r2}. 
[Fig. 3 (a) in the main text is the combination of  Figs.~\ref{fig:plot_r2} (a) and (b).]
In both plots, the location of $t_\updelta=0.5$ with $r_2=0$, the parameter set used in Fig. 2 (b) in the main text, is indicated by an open square, and the location of $t_\updelta =0.5$ with $r_2=-0.2$, the parameter set used in Fig. 3 (b) in the main text, 
is indicated by a filled circle.  

\begin{figure}[h]
\begin{center}
\includegraphics[width=1\columnwidth, clip]{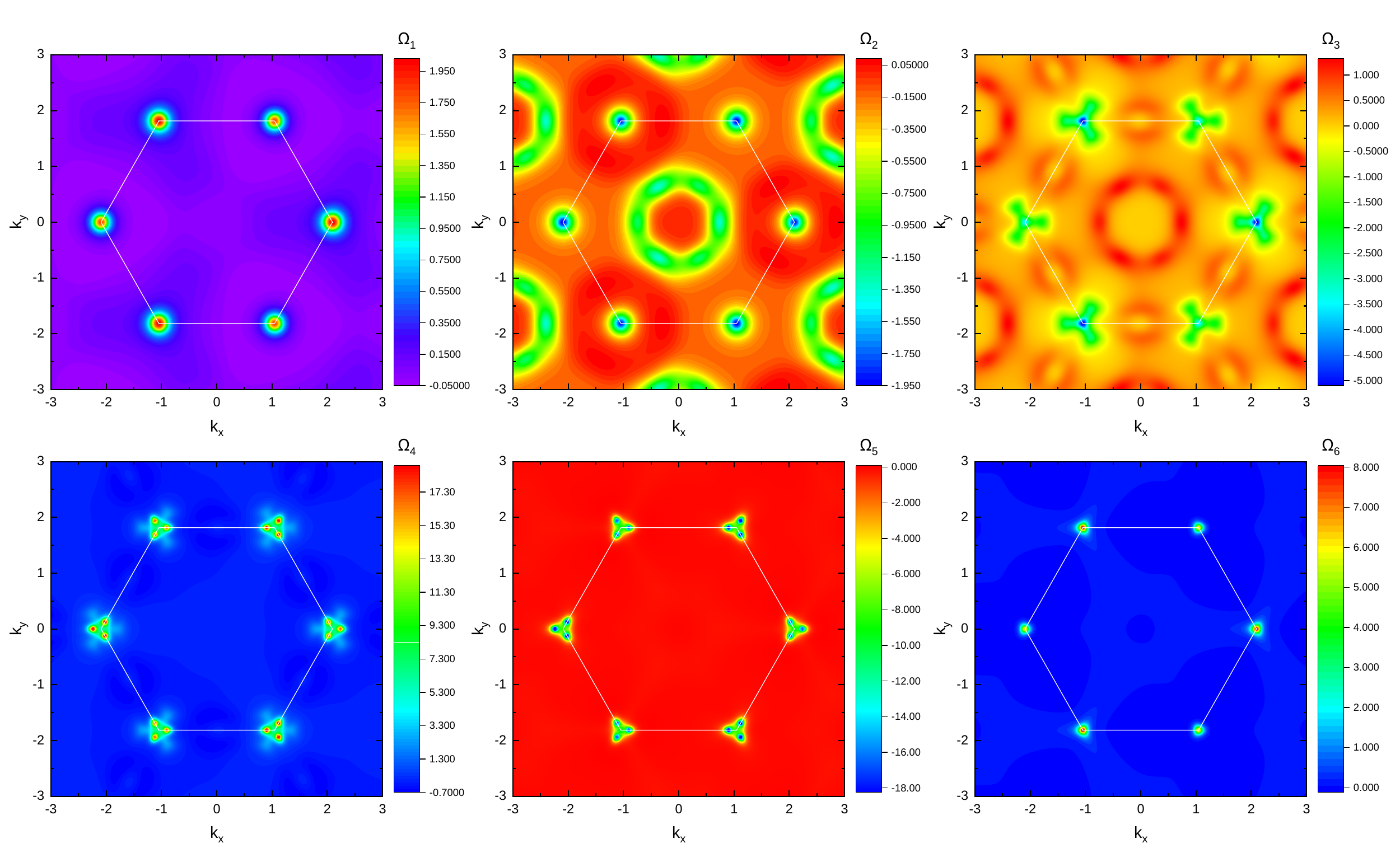}
\caption{Berry curvature $\Omega_i$ of non-interacting model as a function of two-dimensional momentum $(k_x,k_y)$. 
Same as Fig.~\ref{fig:berry_td05} but with $t_\updelta=0.5$ and $r_2=-0.2$. 
}
\label{fig:berry_R105R2m02}
\end{center}
\end{figure}

The Berry curvature of non-interacting electrons with $t_\updelta=0.5$ and $r_2=-0.2$ with $\lambda=0.2$ 
is presented  in Fig.~\ref{fig:berry_R105R2m02}. 

\subsection{$\{xy,x^2-y^2\}$ model}

Here, we show that a two-orbital model involving $xy$ and $x^2-y^2$ orbitals, the $\{xy,x^2-y^2\}$ model, 
has the same form as the $\{yz,xz\}$ model. 
In addition to $t_\uppi$ and $t_\updelta$, we introduce 
$t_\upsigma$, corresponding to $(dd\upsigma)$ in Ref.~\cite{Slater1954Sup}. 
Using these three parameters, the hopping matrices are given by 
\begin{eqnarray}
&\hat t_{\bf 1 \, 2} =  \frac{1}{4}
\left[
\begin{matrix}
4 t_\uppi  & 0 \\
0 & 3 t_\upsigma + t_\updelta 
\end{matrix}
\right],& \nonumber \\ 
&\hat t_{\bf 1 \, 3} = \frac{1}{16} 
\left[
\begin{matrix}
9 t_\upsigma+ 4 t_\uppi + 3 t_\updelta & - \sqrt{3} (3 t_\upsigma - 4 t_\uppi + t_\updelta) \\
- \sqrt{3} (3 t_\upsigma - 4 t_\uppi + t_\updelta) & 3 t_\upsigma+ 12 t_\uppi + t_\updelta
\end{matrix}
\right],& \\ 
&\hat t_{\bf 2 \, 3} = \frac{1}{4} 
\left[
\begin{matrix}
9 t_\upsigma+ 4 t_\uppi + 3 t_\updelta &  \sqrt{3} (3 t_\upsigma - 4 t_\uppi + t_\updelta) \\
 \sqrt{3} (3 t_\upsigma - 4 t_\uppi + t_\updelta) & 3 t_\upsigma+ 12 t_\uppi + t_\updelta
\end{matrix}
\right].& \nonumber
\end{eqnarray}

\begin{figure}
\begin{center}
\includegraphics[width=0.3\columnwidth, clip]{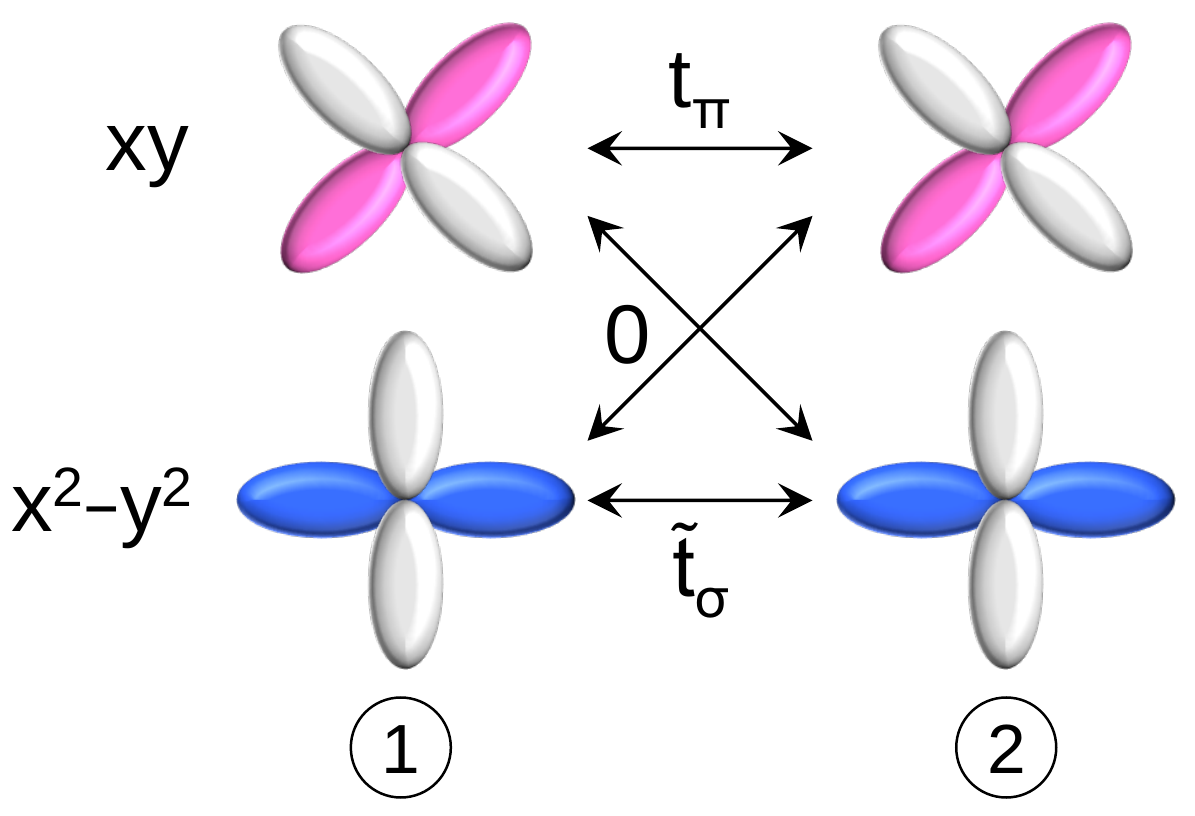}
\caption{Hopping matrix for the $\{xy,x^2-y^2\}$ model. 
$xy (x^2-y^2)$ orbitals between site 1 and site 2 are hybridized via diagonal hopping $t_{\uppi} (\tilde t_{\upsigma})$. 
$\tilde t_{\upsigma}$ is a linear combination of $t_{\updelta}$ and $\upsigma$-bonding $t_{\upsigma}$. 
Other hopping integrals between site 2 and site 3 and  between site 1 and site 3 are obtained via the Slater rule \cite{Slater1954Sup}. 
}
\label{fig:cartoon2}
\end{center}
\end{figure}

Defining $\tilde t_\upsigma = \frac{1}{4}(3 t_\upsigma + t_\updelta)$, 
these matrices are simplified as 
\begin{eqnarray}
&\hat t_{\bf 1 \, 2} =  
\left[
\begin{matrix}
t_\uppi  & 0 \\
0 & \tilde t_\upsigma 
\end{matrix}
\right],& \nonumber \\ 
&\hat t_{\bf 1 \, 3} = \frac{1}{4} 
\left[
\begin{matrix}
3 \tilde t_\upsigma+ t_\uppi & - \sqrt{3} (\tilde t_\upsigma - t_\uppi ) \\
- \sqrt{3} (\tilde t_\upsigma - t_\uppi ) &\tilde t_\upsigma+ 3 t_\uppi 
\end{matrix}
\right],& \\ 
&\hat t_{\bf 2 \, 3} = \frac{1}{4} 
\left[
\begin{matrix}
3 \tilde t_\upsigma+ t_\uppi &  \sqrt{3} (\tilde t_\upsigma - t_\uppi ) \\
 \sqrt{3} (\tilde t_\upsigma - t_\uppi ) &\tilde t_\upsigma+ 3 t_\uppi 
\end{matrix}
\right].& \nonumber 
\end{eqnarray}
A hopping matrix between site 1 and site 2 is schematically shown in Fig.~\ref{fig:cartoon2}. 
One notices that these matrices are identical to the hopping matrices for $\{yz,xz\}$ 
via $(\tilde t_\upsigma, t_\uppi) \rightarrow (t_\uppi, t_\updelta)$ with the change in sign in the off-diagonal elements. 

Using the eigenstate of angular momentum 
$l_z=\pm 2$ for $l=2$, 
the orbitals $|xy\rangle$ and  $|x^2-y^2\rangle$ are expressed as 
$|xy \rangle =- \frac{\rm i}{\sqrt{2}} (|2 \rangle + |-2\rangle)$ 
and $|x^2-y^2 \rangle = \frac{1}{\sqrt{2}} (|2 \rangle - |-2\rangle)$, respectively.   
Using these relations, the SOC term, $\lambda \vec l \cdot \vec s$, in the $\{xy,x^2-y^2\}$ subset is written as 
\begin{eqnarray}
H_{\rm soc} =
\lambda  \sum_{{\bf r} \, \sigma} 
\Bigl( {\rm i} \sigma_{\sigma \sigma}^z  c_{{\bf r} a \sigma}^\dag c_{{\bf r} b \sigma} + {\rm H.c.} \Bigr) .  
\label{eq:soc2}
\end{eqnarray}
This form is also identical to the SOC in the $\{yz,xz\}$ model with a change in prefactor 
from $\frac{1}{2} \lambda$ to $\lambda$.  

\section{Many-body effects}

This section discusses many-body effects in our kagome model. 

\subsection{Effective Hamiltonian}

In multi-orbital systems, the local Coulomb interactions consist of several terms. 
This is simplified in a spin polarized state, where interaction terms contain the single spin component as 
$H_{\rm U}=U \sum_{\bf r} n_{{\bf r} a \uparrow} n_{{\bf r} b \uparrow} 
+ V \sum_{\langle {\bf r} {\bf r}' \rangle} \sum_{\alpha \beta} n_{{\bf r} \alpha \uparrow} n_{{\bf r}' \beta \uparrow}$. 
Here $U$ is the effective Coulomb interaction given by $U = U'-J$ 
with the interorbital Coulomb repulsion $U'$ and the interorbital exchange interaction $J$.
In general, there are other terms, such as the intraorbital Coulomb repulsion $U_0$ and the intraorbital pair transfer $J'$. 
All these parameters are needed to examine the relative stability between a spin polarized ferromagnetic states and others, 
such as antiferromagnetic states and a spin unpolarized state. 

Now we focus on the highest-energy up-spin band, which has the Chern number ${\cal C}_1=1$. 
The kinetic term is given by $\sum_{\rm{\bf k}} \varepsilon_{1 {\rm \bf k}} \psi^\dag_{1 \rm{\bf k}} \psi_{1 \rm{\bf k}}$, 
where $\psi^{(\dag)}_{1 {\rm \bf k}}$ is an annihilation (creation) operator of an electron on the highest-energy band at momentum $\rm{\bf k}$ and energy $\varepsilon_{1 {\rm \bf k}}$. 
$\psi^{(\dag)}_{1 {\rm \bf k}}$ and original electron operators $c^{(\dag)}_{{\rm \bf k} \alpha \uparrow}=\frac{1}{\sqrt{N_{\rm uc}}}
\sum_{\rm \bf r} 
\exp(- {\rm i} {\rm \bf k} \cdot {\rm \bf r})
c^{(\dag)}_{{\rm \bf r} \alpha \uparrow}$
are related by the unitary transformation 
as $c^{(\dag)}_{{\rm \bf k} \alpha \uparrow} = U^{(*)}_{\alpha n} ({\rm \bf k}) \psi^{(\dag)}_{n {\rm \bf k}}$,  
where 
$N_{\rm uc}$ is the total number of unit cells and $n$ is the band index.  
We apply this unitary transformation to $H_{\rm U}$ to obtain effective interactions in the highest band, i.e., $n=1$. 
Since the full expression is rather lengthy, we provide only one example, 
local Coulomb interaction $U n_{\rm{\bf r} a \uparrow} n_{\rm{\bf r} b \uparrow}$. 
This term becomes 
\begin{eqnarray}
\frac{1}{N_{\rm uc}} U \hspace{-0.5em} \sum_{\rm{\bf k}_1,\rm{\bf k}_2,\rm{\bf k}_3,\rm{\bf k}_4} \hspace{-0.5em}
U^*_{a 1} ({\rm{\bf k}}_1) U^*_{b 1} ({\rm{\bf k}}_2) U_{b 1} ({\rm{\bf k}}_3) U_{a 1} ({\rm{\bf k}}_4) \,
\psi^\dag_{1 {\rm \bf k}_1} \psi^\dag_{1{\rm \bf k}_2} \psi_{1{\rm \bf k}_3} \psi_{1{\rm \bf k}_4}
\updelta_{\rm{\bf k}_1+\rm{\bf k}_2-\rm{\bf k}_3-\rm{\bf k}_4}, 
\end{eqnarray}
where 
$\rm{\bf k}_1+\rm{\bf k}_2-\rm{\bf k}_3-\rm{\bf k}_4$ in the $\updelta$ function implies 
$\rm{\bf k}_1+\rm{\bf k}_2-\rm{\bf k}_3-\rm{\bf k}_4$ modulo 
$\rm{\bf b}_1$ and $\rm{\bf b}_2$, 
with $\rm{\bf b}_{1,2}$ being two reciprocal lattice vectors. 
Using lattice translation vectors $\rm{\bf a}_1 = (2,0)$  and $\rm{\bf a}_2=(1,\sqrt{3})$, 
where the distance between nearest-neighbor sites is taken as the unit of length  (see Fig. 1 in the main text), 
the reciprocal lattice vectors are given by ${\rm \bf b}_1 = (\uppi/,\uppi/\sqrt{3})$  and ${\rm \bf b}_2=(0,2\uppi/\sqrt{3})$ (See Fig.~\ref{fig:discrete}). 
The effective Hamiltonian is thus constructed as 
\begin{eqnarray}
H_{\rm eff} = \sum_{\rm{\bf k}} \varepsilon_{1 {\rm \bf k}} \psi^\dag_{1{\rm \bf k}} \psi_{1{\rm \bf k}}
+\frac{1}{N_{\rm uc}} \sum_{{\rm{\bf k}}_1,{\rm{\bf k}}_2, {\rm{\bf k}}_3,{\rm{\bf k}}_4} 
u(\rm{\bf k}_1,\rm{\bf k}_2,\rm{\bf k}_3,\rm{\bf k}_4) \,
\psi^\dag_{1{\rm \bf k}_1} \psi^\dag_{1{\rm \bf k}_2} \psi_{1{\rm \bf k}_3} \psi_{1{\rm \bf k}_4}
\updelta_{\rm{\bf k}_1+\rm{\bf k}_2-\rm{\bf k}_3-\rm{\bf k}_4}. 
\end{eqnarray}
Here, $u(\rm{\bf k}_1,\rm{\bf k}_2,\rm{\bf k}_3,\rm{\bf k}_4) $ involves $U$ and $V$ with appropriate phase factors.

\subsection{Exact diagonalization}

\begin{figure}
\begin{center}
\includegraphics[width=0.4\columnwidth, clip]{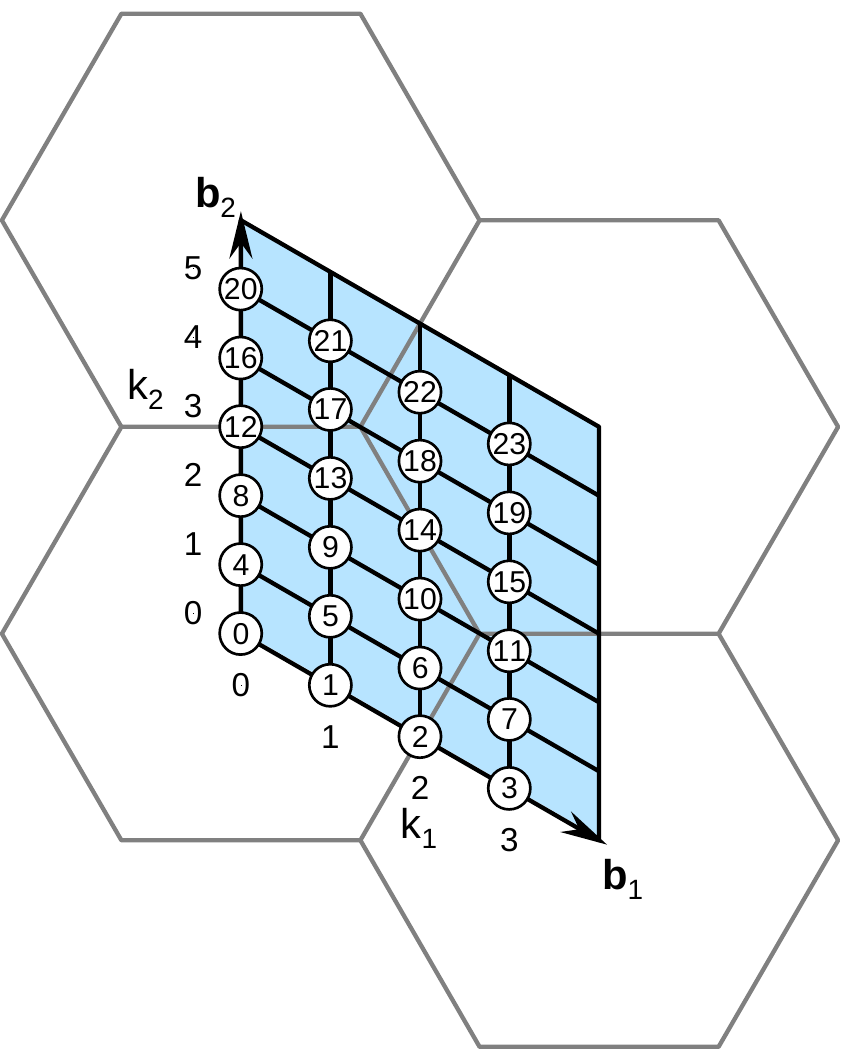}
\caption{Momentum discretization used in exact diagonalization calculations for the interacting model. 
Two reciprocal lattice vectors, ${\rm\bf a}_{1}$ and ${\rm\bf a}_{2}$, are divided by 
$N_1=4$ and $N_2=6$, respectively. 
A number in a circle on each momentum is $k_1 + N_1 \times k_2$. 
}
\label{fig:discrete}
\end{center}
\end{figure}

The effective Hamiltonian is diagonalized in momentum space. 
In this work, we consider a rhombus spanned by the two reciprocal lattice vectors and divide it into $N_1 \times N_2$ patches. 
This corresponds to having $N_{1(2)}$ unit cells along the ${\rm\bf a}_{1(2)}$ direction, so $N_{\rm uc}=N_1 \times N_2$. 
Discretized momenta are now expressed as ${\rm \bf k} = {\rm \bf b}_1 k_1/N_1 + {\rm \bf b}_2 k_2/N_2$. 
We simply label discretized momenta using $(k_1,k_2)$. 
Figure~\ref{fig:discrete} depicts the momentum discretization used in this study  with $N_1=4, N_2=6$ and $N_{\rm uc}=24$

In this study, we consider a fractional filling $\nu=1/3$. 
With $N_{\rm uc}=24$, the number of electrons on the highest band is $N_{\rm e}=N_{\rm uc}/3=8$. 
In this case, the size of the Hilbert space is $_{24} C_8=735,471$. 
Due to the translational symmetry and the momentum conservation of many-body interaction terms, 
$H_{\rm eff}$ is subdiagonalized according to the total momentum, 
${\rm \bf k}_{\rm tot}=  \sum_{\rm{\bf k \in l}} {\rm \bf k}$ 
modulo 
$\rm{\bf b}_1$ and $\rm{\bf b}_2$. 
Using the same notation as the momentum discretization $(k_1,k_2)$, each momentum sector has roughly $735,471/24 \approx 30,700$ states. 
The number of states of each momentum sector for our system is summarized in Table~\ref{tab:number}.
$30,700\times30,700$ Hamiltonian matrices can be diagonalized efficiently using, for example, ARPACK \cite{ARPACK}. 

\begin{table}
\caption{List of the number of states at each sector $(k_1,k_2)$
for the $\nu=1/3$ filling of 24 site cluster with $N_1=4$, $N_2=6$, and $N_{\rm e}=8$.}
\label{tab:number}
\begin{center}
\begin{tabular}{c c| cccc}
&& \multicolumn{4}{c}{ $k_1$ } \\
&& 0 & 1 & 2 & 3 \\
\hline
{\multirow{6}{*}{ $k_2$ }}& 0 \, & \, 30,709 \, & \, 30,624 \, & \, 30,704 \, & \, 30,624 \, \\
{}&1 \, & 30,624 & 30,624 & 30,624 & 30,624 \\
{}&2 \, & 30,709 & 30,624 & 30,704 & 30,624 \\
{}&3 \, & 30,624 & 30,624 & 30,624 & 30,624 \\
{}&4 \, & 30,709 & 30,624 & 30,704 & 30,624 \\
{}&5 \, & 30,624 & 30,624 & 30,624 & 30,624
\end{tabular}
\end{center}
\end{table}

\subsection{Many-body Chern number}

Many-body Chern number of the ground state $(k_1,k_2)$ is computed via 
${\cal C}_{(k_1,k_2)} = \frac{1}{2 \uppi} \! \int_0^{2 \uppi} \! d \theta_1 \int_0^{2 \uppi} \! d \theta_2 F_{(k_1,k_2)} (\theta_1, \theta_2)$ \cite{Niu1985Sup}
where $F(\theta_1, \theta_2)$ is the Berry curvature given by
\begin{eqnarray}
F_{(k_1,k_2)}(\theta_1, \theta_2)= {\rm Im} \biggl\{
\bigg\langle \frac{\partial \Phi_{(k_1,k_2)}}{\partial \theta_2} \bigg| \frac{\partial \Phi_{(k_1,k_2)}}{\partial \theta_1} \bigg\rangle
-\bigg\langle \frac{\partial \Phi_{(k_1,k_2)}}{\partial \theta_1} \bigg| \frac{\partial \Phi_{(k_1,k_2)}}{\partial \theta_2} \bigg\rangle
\biggr\},  
\label{eq:berry}
\end{eqnarray}
and $| \Phi_{(k_1,k_2)} \rangle$ is the many-body wave function constructed using single-particle wave functions 
$\psi ({\rm{\bf r}})$ after the Fourier transformation to momentum space. 
The momentum index $(k_1,k_2)$ will be omitted in the following discussion for simplicity. 

Partial derivative of a wave function with respect to $\theta_j$ is approximated by the finite difference 
as $|\partial \Phi/\partial \theta \rangle \approx \frac{1}{|\updelta {\boldsymbol \theta}|} 
[| \Phi ({\boldsymbol \theta} + \updelta {\boldsymbol \theta}) \rangle - | \Phi ({\boldsymbol \theta}) \rangle]$. 
Here, the vector notation is used for ${\boldsymbol \theta} = (\theta_1,\theta_2)$, and $\updelta {\boldsymbol \theta} = (\updelta \theta_1,0)$ or $(0,\updelta \theta_2)$. 
Then, it is required to compute a product of two wave functions as 
$\langle \Phi ({\boldsymbol \theta})|\Phi ({\boldsymbol \theta}') \rangle$ with $\boldsymbol \theta \ne \boldsymbol \theta'$. 

Since we are using a multiorbital model projected onto the flat band, a special care is needed. 
By diagonalizing a many-body Hamiltonian, the ground state in the momentum sector $(k_1,k_2)$ is given by 
\begin{eqnarray}
| \Phi ({\boldsymbol \theta}) \rangle = \sum_l V_{1 l} ({\boldsymbol \theta}) |\varphi_l ({\boldsymbol \theta}) \rangle,
\end{eqnarray}
where $V_{1l} (\boldmath \theta)$ is the $(1,l)$ component of the unitary matrix, which diagonalizes the many-body Hamiltonian in the momentum sector $(k_1,k_2)$, 
and $1$ means the ground state. 
$|\varphi_l ({\boldsymbol \theta}) \rangle$ form the basis set to describe the many-body Hilbert space. 
Specifically, this is given by 
\begin{eqnarray}
|\varphi_l ({\boldsymbol \theta}) \rangle = \prod_{{\rm \bf k} \in l} \psi^\dag_{1 {\rm \bf k}+ \updelta {\rm \bf k}} |0\rangle. 
\end{eqnarray}
Here, $\psi^\dag_{1{\rm \bf k}+ \updelta {\rm \bf k}}$ is a creation operator of an electron at the highest flat band at momentum ${\rm \bf k}+ \updelta {\rm \bf k}$, 
and $|0\rangle$ is the vacuum. 
Note that $V$ matrices depend on the total momentum $(k_1,k_2)$, but $U$ matrices do not because it is from diagonalising a single particle Hamiltonian. 
The product over $\rm \bf k$ is limited to a set of momenta specified by the index $l$. 
The momentum shift $\updelta {\rm \bf k}$ and the twist phase $\boldsymbol \theta$ are related via 
$\updelta {\rm \bf k} = {\rm \bf b}_1 \theta_1 /2 \uppi N_1 + {\rm \bf b}_2 \theta_2 /2 \uppi N_2$. 
Recalling that a creation operator $\psi^\dag_{1{\rm \bf k}}$ is rewritten by creation operators of original electrons $c^\dag_{{\rm \bf k} \alpha \uparrow}$ as 
$\psi^{\dag}_{1{\rm \bf k}} = \sum_{\alpha} U_{\alpha 1} (\rm{\bf k})  c^{\dag}_{\rm{\bf k} \alpha \uparrow} $, one finds 
\begin{eqnarray}
|\varphi_l ({\boldsymbol \theta}) \rangle = \prod_{{\rm \bf k} \in l} 
\Biggl[ \sum_\alpha U_{\alpha 1} ({\rm \bf k} + \updelta {\rm \bf k})  c^{\dag}_{{\rm \bf k} + \updelta {\rm \bf k} \alpha \uparrow} \Biggr] |0\rangle. 
\end{eqnarray}
Using this, $\langle \Phi ({\boldsymbol \theta})|\Phi ({\boldsymbol \theta}') \rangle$ is rewritten as 
\begin{eqnarray}
\langle \Phi ({\boldsymbol \theta}) 
|\Phi ({\boldsymbol \theta}') \rangle \!\!&=&\!\! \langle 0| \sum_l V_{1l}^* ({\boldsymbol \theta}) 
\prod_{{\rm \bf k} \in l} \Biggl[ \sum_\alpha U_{\alpha 1}^* ({\rm \bf k} + \updelta {\rm \bf k})  c_{{\rm \bf k}+\updelta {\rm \bf k} \alpha \uparrow} \Biggr] \nonumber \\
&&\!\! \times \sum_{l'} V_{1l'} ({\boldsymbol \theta}')
\prod_{{\rm \bf k}' \in l'} \Biggl[ \sum_\beta U_{\beta 1} ({\rm \bf k}' + \updelta {\rm \bf k}')  
c^{\dag}_{{\rm \bf k}'+\updelta {\rm \bf k}' \beta \uparrow} \Biggr] |0\rangle. 
\end{eqnarray}
Considering small $|\boldsymbol \theta - \boldsymbol \theta'|$, 
because $ \boldsymbol \theta'=\boldsymbol \theta + \updelta \boldsymbol \theta$, 
$\langle 0| c_{{\rm \bf k}+\updelta {\rm \bf k} \alpha \uparrow} c^\dag_{{\rm \bf k}'+\updelta {\rm \bf k}' \beta \uparrow} |0 \rangle$ is approximated 
to be $\updelta_{\alpha \beta} \updelta_{{\rm \bf k} {\rm \bf k}'}$. 
This leads to a simple expression 
\begin{eqnarray}
\langle \Phi ({\boldsymbol \theta}) 
|\Phi ({\boldsymbol \theta}') \rangle \approx \sum_l V_{1l}^* ({\boldsymbol \theta}) V_{1l} ({\boldsymbol \theta}') 
\prod_{{\rm \bf k} \in l} \Biggl[ \sum_\alpha U_{\alpha 1}^* ({\rm \bf k} + \updelta {\rm \bf k})  U_{\alpha 1} ({\rm \bf k} + \updelta {\rm \bf k}') \Biggr]. 
\end{eqnarray}

This is used to compute products of two wave functions in Eq.~(\ref{eq:berry}). 
For the integral over two boundary phases, we divide the boundary phase unit cell into $20 \times 20$ meshes, numerically evaluate the partial derivative 
of the many-body wave function on each placket using a technique proposed in Ref.~\cite{Fukui2005}, and sum up all the quantities. 



\end{document}